\documentclass[11pt]{article}

\usepackage[a4paper,margin=2.8cm,bottom=3.8cm]{geometry}

\usepackage{graphicx}
\usepackage{amssymb}
\usepackage{amsmath}
\usepackage{amsthm}
\usepackage{mathrsfs}
\usepackage{url}
\usepackage{color}
\usepackage{longtable}
\usepackage{rotating}
\usepackage{authblk}
\usepackage{multirow}
\usepackage[authoryear]{natbib}

\definecolor{red}{rgb}{1,0,0}

\begin{document}

\title{Counterparty Credit Limits: The Impact of a Risk-Mitigation Measure on Everyday Trading}

\author[1,2]{Martin D. Gould\thanks{Corresponding author. Email: gouldm@maths.ox.ac.uk.}
\thanks{Martin D. Gould completed part of this work while as an academic visitor at the Humboldt-Universit\"{a}t zu Berlin.}}
\author[3]{Nikolaus Hautsch}
\author[1]{Sam D. Howison}
\author[1,4]{Mason A. Porter}
\affil[1]{Mathematical Institute, University of Oxford, Oxford OX2 6GG, UK}
\affil[2]{CFM--Imperial Institute of Quantitative Finance, Department of Mathematics, Imperial College, London SW7 2AZ, UK}
\affil[3]{Department of Statistics and Operations Research, Research Platform ``Data Science @ Uni Vienna'', University of Vienna, A-1090 Vienna, Austria}
\affil[4]{Department of Mathematics, University of California, Los Angeles, CA 90095, USA}
\date{}

\maketitle

\begin{abstract}

A counterparty credit limit (CCL) is a limit that is imposed by a financial institution to cap its maximum possible exposure to a specified counterparty. CCLs help institutions to mitigate counterparty credit risk via selective diversification of their exposures. In this paper, we analyze how CCLs impact the prices that institutions pay for their trades during everyday trading. We study a high-quality data set from a large electronic trading platform in the foreign exchange spot market, which enables institutions to apply CCLs. We find empirically that CCLs had little impact on the vast majority of trades in this data. We also study the impact of CCLs using a new model of trading. By simulating our model with different underlying CCL networks, we highlight that CCLs can have a major impact in some situations.

\end{abstract}




\section{Introduction}\label{sec:introduction}

The international financial crisis of 2008 underlines the vital importance of understanding counterparty credit risk. The collapse of Lehman Brothers and the ensuing defaults and near-defaults by AIG, Bear Stearns, Fannie Mae, Freddie Mac, Merrill Lynch, the Icelandic banks, and the Royal Bank of Scotland demonstrated how the complex and highly interconnected nature of the modern financial ecosystem can cause counterparty failures to propagate rapidly between institutions and can thereby amplify their severity \citep{May:2008complex}. Consequently, it is extremely important to assess and implement measures to mitigate this risk. 

One possible mitigation measure, which has been implemented by several major trading platforms in the foreign exchange (FX) spot market, is the use of \emph{counterparty credit limits} (CCLs). A CCL is a limit that is imposed by a financial institution to cap its maximum possible exposure to a specified counterparty. With CCLs in place, institutions can access only the subset of trading opportunities that are offered by counterparties with whom they possess sufficient bilateral credit.

Limiting exposures in this way creates both benefits and drawbacks for individual financial institutions. On one hand, CCLs can protect an institution from suffering large losses as a result of a counterparty default. On the other hand, CCLs restrict the set of trading opportunities that an institution can access. This can result in an institution trading at a less favourable price than would be the case without such restrictions. Importantly, this restriction applies at all times, rather than only during periods of market-wide stress (when the default-protection benefit of CCLs is most useful).

One aim of the present paper is to assess how CCLs affect the prices that institutions pay for their trades during everyday trading. To study this question, we use an unusually rich data set that describes all trading activity for three liquid currency pairs on Hotspot FX during all of May and June 2010. Hotspot FX\footnote{In February 2017, Hotspot FX was acquired by Cboe Global Markets and rebranded as ``Cboe FX Markets''. At the time of writing, Cboe FX Markets remains a major electronic FX spot trading platform with a large market share \citep{CboeVolumes}.} is a large electronic trading platform in the FX spot market that allows institutions to apply CCLs. The Hotspot FX data enables us to measure how CCLs impact the prices that individual institutions pay for their trades. Because the period from May to June in 2010 was a relatively calm period with no major institutional failures, we are able to study the impact of CCLs during `normal' trading. To the best of our knowledge, our study is the first one to investigate this topic.

We introduce the notion of the ``skipping cost'' of a trade to measure the additional cost that an institution bears from the application of CCLs. In our data set, more than half of the trades have a skipping cost of zero, and the mean skipping cost is less than half of a basis point. We do identify a handful of trades with large skipping costs, but we argue that the existence of such trades is a natural consequence of the heterogeneity in the types and sizes of institutions that trade on Hotspot FX. We also find that the realized volatility of trade prices is very similar to the corresponding realized volatility in the platform-wide best quotes.

Overall, our empirical results suggest that CCLs had little impact on the vast majority of the trades that we study. However, the empirical study of historical data does not provide insight into how these results may change if institutions make substantial modifications to their CCLs. We therefore complement our empirical analysis by investigating a second question: How does CCL network structure affect the prices that institutions pay for their trades? To do this, we simulate an agent-based model of trade in which institutions assign CCLs to each other. In contrast to our empirical analysis, in which the CCL network is fixed and unobservable to us, our model allows us to investigate how varying the CCL network can affect the prices of trades.

Our simulations illustrate that when the CCL network is dense (in the sense that most institutions can access most trading opportunities), CCLs have very little impact on the prices of trades. However, as the edge density falls, we observe that the skipping costs of trades rise sharply and that the trade-price volatility rises sharply, whereas the quote-price volatility remains almost constant. We also observe that network topology has noticeable effects on our results. For example, for a given edge density, the mean skipping cost is markedly higher for a core--periphery network than it is for an Erd\H{o}s--R\'{e}nyi network.

Together, our empirical and simulation results paint a complex picture about the impact of CCLs, raising important questions for policy makers and regulators. Both sets of results illustrate that the application of CCLs does not necessarily result in many institutions paying large skipping costs for their trades. Therefore, CCLs may provide market participants with the benefits of selective diversification without causing them to incur large additional costs during  everyday trading. However, our model also illustrates how an aggressive application of CCLs can lead to much larger skipping costs and can also create large jumps in the trade-price series, even when the quote-price series remains relatively stable. We thus argue that understanding and monitoring how institutions set and adjust their CCLs is vital for regulators when assessing how the implementation of CCLs may impact market stability.

Our paper proceeds as follows. In Section~\ref{sec:lit_rev}, we give an introduction to counterparty credit risk and discuss a variety of relevant literature. In Section~\ref{sec:ccls}, we describe the CCL mechanism in detail and discuss how CCLs are currently implemented by several large electronic trading platforms in the FX spot market. We present our empirical results in Section~\ref{sec:empirics}. In Section~\ref{sec:model}, we introduce and study our model of trading with CCLs. In Section~\ref{sec:conclusion}, we discuss our results and present our conclusions. In the appendices, we give a detailed description of our data and describe our methodology for estimating realized volatility.


\section{Literature Review}\label{sec:lit_rev}

\emph{Counterparty credit risk} is the risk that one or more of a financial institution's counterparties default on their agreed obligations. For a detailed survey, see~\citet{Gregory:2010counterparty}. In this section, we review a variety of studies to provide context for our work.

\citet{Jarrow:2001counterparty} observed that financial institutions face significant counterparty credit risk whenever their exposures are concentrated in a small number of counterparties, because the default of any such counterparty is likely to cause severe financial distress. In real financial markets, most financial institutions trade with a wide range of different counterparties, creating a network of interconnected credit relationships. As noted in \citet{Stiglitz:2010risk} and \citet{Roukny:2013default}, these relationships create both benefits and drawbacks for counterparty credit risk. On one hand, network connections diversify financial institutions' risk exposures; on the other hand, they create contagion channels through which shocks can spread.

Many studies (see, e.g., \citet{Anand:2012rollover, Bardoscia:2017pathways, Gai:2011complexity, Gandy:2017bayesian, Giesecke:2004cyclical, Hautsch:2015financial, Jorion:2009credit, May:2008complex, Battiston:2012default, Battiston:2012liaisons}) have illustrated how contagion channels can cause default cascades and can thereby lead to systemic risk. See \citet{Glasserman:2016contagion} and \citet{Jackson:2020systemic} for surveys. A smaller number of studies have focused on how the topology of financial networks impacts such dynamics. \citet{Roukny:2013default} compared the size of default cascades in Erd\H{o}s--R\'{e}nyi networks and in networks with heavy-tailed degree distributions, and reported that network topology strongly impacts the probability and size of default cascades. From the perspective of systemic risk, they concluded that no single market topology is always better than all others. \citet{Luu:2018collateral} examined how network topology affects the dynamics of collateral (and the consequent systemic risk) in the presence of rehypothecation. They observed starkly different dynamics for different network topologies. \citet{Weber:2017joint} introduced a multi-factor model of how bankruptcy costs, fire sales, and cross-holdings impact systemic risk, and simulated their model using both Erd\H{o}s--R\'{e}nyi networks and core--periphery networks.\footnote{We use the same two classes of networks in our simulations of our model of trade via CCLs in Section \ref{sec:model}.} They observed qualitatively similar results for both classes of networks (albeit with quantitative differences that depend on the precise edge placements in the networks). These studies all underline the potential severity of counterparty credit risk in the modern financial ecosystem and thereby provide strong motivation for exploring safeguards against it. 

One possible approach to mitigating counterparty credit risk is to novate trade via a central counterparty (CCP). See \citet{Norman:2011risk} and \citet{Rehlon:2013central} for detailed discussions. The role of a CCP is to guarantee the obligations that arise from all contracts that are agreed between two counterparties. If one counterparty fails, the other is protected via the resources and default-management procedures of the CCP. During the past decade, several prominent regulatory bodies (see, e.g., \citet{Basel:2013supervisory} and \citet{CRMPG:2005toward}) have argued that CCPs are an effective tool for mitigating counterparty credit risk. However, several authors have noted that trade novation via a CCP also entails drawbacks. \citet{Pirrong:2012clearing} argued that CCP novation does not reduce the aggregate counterparty credit risk across all institutions; instead, it concentrates all such risk into the CCP, which thus becomes a single point of failure with systemic importance. \citet{Biais:2012clearing} noted that although CCPs allow mutualization of the idiosyncratic risk faced by individual institutions, they cannot provide protection against the aggregate risk that affects all institutions. \citet{Menkveld:2015crowded} reported that the risk-management methodologies that are implemented by CCPs can greatly underestimate the probability of clustered defaults, which place severe stress on a CCP. \citet{Koeppl:2013limits} noted that CCPs generate a moral hazard by reducing the incentives for individual institutions to assess the creditworthiness of their trading counterparties. Given the historical failures of several CCPs in a wide variety of asset classes \citep{Gregory:2010counterparty}, concerns about whether CCPs truly mitigate counterparty credit risk, or simply repackage it, seem to be well-founded.

Another possible approach to mitigating counterparty credit risk is to apply a credit valuation adjustment (CVA). See \citet{Brigo:2013counterparty} for a detailed discussion. In this framework, an institution adjusts the price that it offers another institution to account for the risk of trading with it. \citet{Brigo:2013counterparty} introduced a method for calculating a CVA by pricing a contingent claim whose payoff is triggered by the default of the given counterparty, such that the resulting net loss is $0$. Recently, \citet{Barucca:2020network} and \citet{Banerjee:2020pricing} extended this approach by deriving formulas for CVAs in a network of financial institutions that are interconnected via creditor and debtor relationships and correlated via similarities in their balance sheets. However, even with these extensions, CVA suffers from important practical and theoretical drawbacks. From a practical perspective, it is not possible for institutions to use CVAs when trading on an exchange in which many different institutions access the same centralized set of trading opportunities. For example, this is the case in a limit order book (LOB). (See \citet{Gould:2013limit} for a detailed introduction to LOBs.) From a theoretical perspective, \citet{Cesari:2010modelling} noted that calculating even a single CVA requires an institution to estimate a specific counterparty's default probability; this is extremely difficult in practice. They also remarked that CVA estimation provides no insight into how to construct an asset portfolio that gives the required payoff upon a counterparty default, and they noted that constructing such a portfolio is often impossible.

The above weaknesses suggest that neither CCPs nor CVAs are a panacea for counterparty credit risk. Their failure to provide a conclusive solution to the problem is strong motivation for exploring alternative avenues.


\section{Counterparty Credit Limits}\label{sec:ccls}

Consider a financial market that is populated by a set $\Theta=\left\{\theta_1,\theta_2,\ldots\right\}$ of institutions, and suppose that each institution $\theta_i$ assigns a CCL $c_{(i,j)}\geq 0$ to each other institution $\theta_j$. The CCL $c_{(i,j)}$ specifies the maximum level of counterparty credit exposure that $\theta_i$ is willing to extend to $\theta_j$. Such counterparty credit exposures occur in all financial markets in which the agreement and settlement of trades does not occur simultaneously. For example, in the FX spot market, trades that are agreed on day $D$ are settled on day $D+2$, so each trade entails exposure to the counterparty during the period between day $D$ and day $D+2$.

No institution $\theta_i$ can trade with any other institution $\theta_j$ if doing so would make $\theta_i$'s total exposure to $\theta_j$ exceed $c_{(i,j)}$, or would make $\theta_j$'s total exposure to $\theta_i$ exceed $c_{(j,i)}$. The maximum amount that $\theta_i$ and $\theta_j$ can trade is therefore equal to $\min\left(c_{(i,j)},c_{(j,i)}\right)$. We call this amount the \emph{bilateral CCL between $\theta_i$ and $\theta_j$}. Bilateral CCLs determine the subset of trading opportunities that are available to each institution. This subset changes over time according to the relevant institutions' trading activities.

Institutions can use CCLs to mitigate counterparty credit risk by selective diversification of their exposures. If an institution $\theta_i$ perceives another institution $\theta_j$ to be unacceptably likely to default, then $\theta_i$ can ensure that it never trades with $\theta_j$ by setting $c_{(i,j)}= 0$. Alternatively, if $\theta_i$ perceives $\theta_j$ to be extremely unlikely to default, then $\theta_i$ can assign an unlimited amount of credit to $\theta_j$ by setting $c_{(i,j)}= \infty$.

In contrast to trade novation via a CCP, the application of CCLs does not require either (1) a single, centralized clearing node that constitutes a single point of failure for an entire market or (2) that every institution posts collateral.\footnote{Note, however, that the use of CCLs does not exclude the subsequent clearing of trades via a CCP. We return to this discussion in Section~\ref{sec:conclusion}.} In contrast to CVAs, the use of CCLs does not require institutions to estimate the market value of their counterparty credit risk. Instead, CCLs enable all institutions to specify an upper bound for each of their counterparty exposures.

Several major multi-institution electronic trading platforms in the FX spot market offer institutions the ability to implement CCLs. On these platforms, each institution $\theta_i$ privately declares (to the exchange) their CCL $c_{(i,j)}$ for each other institution $\theta_j$. Trades occur via a mechanism that is similar to a standard limit order book (LOB), except that institutions can only conduct transactions that do not violate their bilateral CCLs. More precisely, when an institution $\theta_i$ submits a buy (respectively, sell) market order, the order matches to the highest-priority sell (respectively, buy) limit order that is owned by an institution $\theta_j$ such that neither $c_{(i,j)}$ nor $c_{(j,i)}$ is exceeded by conducting the given trade. We call this market organization a \emph{quasi-centralized LOB} (QCLOB) because different institutions have access to different subsets of the same (otherwise centralized) LOB. For a detailed introduction to QCLOBs, see \citet{Gould:2016quasi}.

Institutions that trade on a QCLOB platform cannot, in general,\footnote{The Reuters and Electronic Broking Services (EBS) platforms offer institutions an additional data feed that, in exchange for a fee, provides snapshots of the global LOB at regular time intervals.} see the state of the global LOB (i.e., the set of all orders that are owned by all market participants on the platform). Instead, each institution sees only the subset of orders that correspond to its own trading opportunities (i.e., that would not violate any of its bilateral CCLs upon execution of a trade) at time $t$. More precisely, for each $j\neq i$, the volume of each separate limit order placed by $\theta_j$ that is visible to $\theta_i$ is reduced (if necessary) so that its size does not exceed the bilateral CCL between $\theta_i$ and $\theta_j$. Each institution therefore views a filtered set of all limit orders on the platform.

As well as viewing their filtered LOB, each institution in a QCLOB can access a \emph{trade-data stream}, which lists the price, time, and direction (buy or sell) of every trade that occurs. All institutions can see all entries in the trade-data stream in real time, irrespective of their bilateral CCLs. Therefore, although institutions in a QCLOB can see only a subset of the trading opportunities that are available to other institutions, they have access to a detailed historical record of all previous trades. 


\section{Empirical Results}\label{sec:empirics}

Our empirical investigation uses a data set that was provided to us by Hotspot~FX. The data set gives all trading activity on the Hotspot~FX platform for the EUR/USD (euro/US dollar), GBP/USD (pounds sterling/US dollar), and EUR/GBP (euro/pounds sterling) currency pairs\footnote{A price for the currency pair XXX/YYY denotes how many units of the \emph{counter currency} YYY are exchanged per unit of the \emph{base currency} XXX.} for the entire months of May and June in 2010. According to the 2010 Triennial Central Bank Survey \citep{BIS:2010triennial}, global trade for these currency pairs constituted about $28\%$, $9\%$, and $3\%$ of the total turnover of the FX market, respectively. The period from May through June in 2010 was a relatively calm period in the FX spot market, with no major institutional failures. This makes it suitable for studying how CCLs impact the prices that institutions pay for their trades during everyday trading. For a detailed description of the Hotspot FX data, see Appendix~A.

\subsection{Skipping Costs}\label{sec:skippingcosts}

We first examine the question of how CCLs impact the prices of individual trades. As we discussed in Section~\ref{sec:ccls}, when an institution $\theta_i$ on Hotspot FX submits a buy (respectively, sell) market order, the order matches to the highest-priority sell (respectively, buy) limit order that is owned by an institution $\theta_j$ such that the bilateral CCL between $\theta_i$ and $\theta_j$ is not violated. Therefore, the price at which a given market order matches is not necessarily the best price that is available to other institutions at that time. The Hotspot FX data enables us to calculate the difference between the price at which each trade occurs and the lowest price among all sell limit orders (respectively, the highest price among all buy limit orders) at the same instant. It therefore enables us to quantify the additional cost (to the institution submitting the market order) as a result of CCLs preventing this institution from accessing a better-priced trading opportunity. We call this additional cost the ``skipping cost''.

For a given currency pair on a given trading day, let $p_k$ denote the price of the $k^{\text{th}}$ trade and let $b_k$ and $a_k$ denote, respectively, the bid- and ask-price in the global LOB immediately before this trade occurs. The \emph{skipping cost} of a trade is 
\begin{equation}\label{eq:rk}r_k = \left\{\begin{array}{ll}
	\displaystyle{p_{k}-a_{k}\,,} & \text{ if the $k^{\text{th}}$ trade is buyer-initiated} \\
	\displaystyle{b_{k}-p_{k}\,,} & \text{ if the $k^{\text{th}}$ trade is seller-initiated\,.}\end{array}\right.
\end{equation}
The sign difference in Equation (\ref{eq:rk}) ensures that every trade has a non-negative skipping cost. In the extreme case in which all institutions always have access to all trading opportunities, all trades occur at the best quotes, so $r_k=0$ for all $k$. In this case, price formation is equivalent to that in a standard LOB.

Because the prices of trades vary across currency pairs and across time, we also normalize each skipping cost by the mid-price $m_k=0.5(a_k+b_k)$ immediately before a trade occurs. Specifically, we calculate the \emph{normalized skipping cost}
\begin{equation}\label{eq:rktilde}
	\tilde{r}_k =\frac{r_k}{m_k}\,,
\end{equation}
which we measure in basis points (where $1$ basis point equals $0.01\%$). We apply this scaling to make $\tilde{r}_k$ independent of the size of the underlying exchange rate; this allows easier comparisons across different currency pairs.

\begin{figure}[!htbp]
\centering
\includegraphics[width=0.49\textwidth]{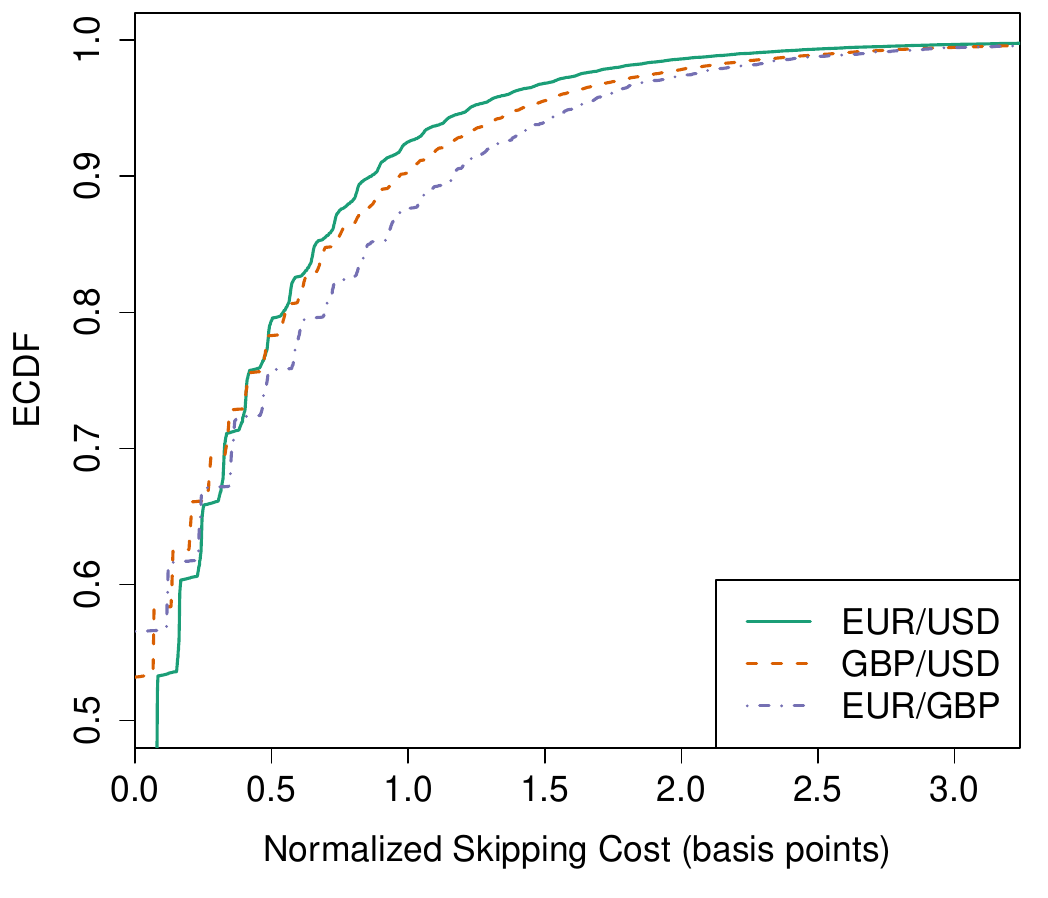}
\includegraphics[width=0.49\textwidth]{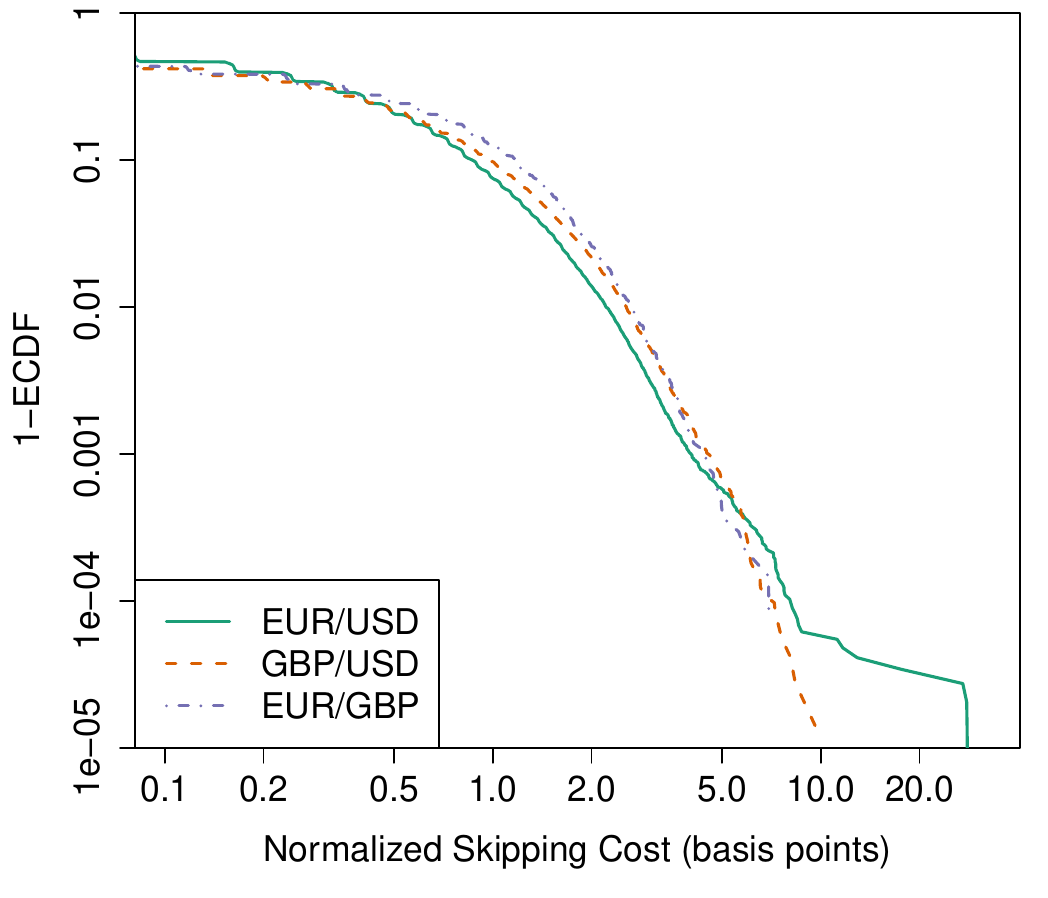}
\caption{(Left) Empirical cumulative density function (ECDF) and (right) log-survival function (i.e., 1 minus the ECDF) for the normalized skipping costs $\tilde{r}_k$.}
\label{fig:Skipping}
\end{figure}

In Figure~\ref{fig:Skipping}, we show the empirical cumulative density functions (ECDFs) of normalized skipping costs $\tilde{r}_k$. More than half of all trades have a normalized skipping cost of $0$, which implies that they occurred at the best price that was available in the global LOB at their time of execution. Up to about the $99.9^\text{th}$ percentile, the distribution of normalized skipping costs has a similar shape for all three currency pairs. Beyond this point, EUR/USD includes a handful of trades with extremely large skipping costs; this does not occur for the other two currency pairs.

In Table~\ref{tab:SkippingSummaryStats}, we give summary statistics about the normalized skipping costs $\tilde{r}_k$. For each of the three currency pairs, the mean normalized skipping cost is about $0.2$ basis points and the standard deviation of normalized skipping costs is about $0.5$ basis points (i.e., $0.005\%$). As with Figure \ref{fig:Skipping}, these results suggest that the statistical properties of normalized skipping costs are similar for each of the three currency pairs.

\begin{table}[!htbp]
\centering
\begin{tabular}{lrrr}
\hline
& EUR/USD & GBP/USD & EUR/GBP \\
\hline
Minimum & $0$ & $0$ & $0$ \\ 
Median & $0$ & $0$ & $0$ \\ 
Maximum & $30.31$ & $9.65$ & $5.62$ \\ 
Mean & $0.19$ & $0.21$ & $0.21$ \\ 
Standard Deviation & $0.46$ & $0.43$ & $0.45$ \\ 
\hline
\end{tabular}
  \caption{Summary statistics for normalized skipping costs $\tilde{r}_k$.}
\label{tab:SkippingSummaryStats}
\end{table}

When considering the raw skipping costs $r_k$ (i.e., without normalization), the mean skipping costs range from about $1.8$ ticks (for EUR/GBP) to about $3.0$ ticks (for GBP/USD). Given that the tick size for each of the three currency pairs is $0.00001$ units of the counter currency (see Appendix~A), these skipping costs correspond to a mean additional cost of about $\pounds 18$ and $\$30$, respectively, for an institution that submits a market order for $1$ million units (which is the modal market order size for each of the three currency pairs). Although these mean skipping costs are relatively small, some trades in our sample have much larger skipping costs. The largest skipping cost that we observe exceeds $30$ basis points, and corresponds to incurring an additional cost of about $\$3630$ when submitting a trade of size $1$ million euros.

As we discussed in Section \ref{sec:ccls}, institutions on Hotspot FX can infer the approximate skipping cost of their trades by comparing their local bid-price or ask-price (which they can observe from the filtered set of limit orders that they observe on the platform) to the prices of recent trades (which they can observe via their trade-data stream). Given that this is the case, why do some institutions perform trades that have extremely large skipping costs during everyday trading? We believe that the answer lies in the fact that Hotspot FX serves a wide variety of institutions, which have varying levels of access to other trading mechanisms. At times when submitting a market order would entail a considerable skipping cost, large institutions would likely instead perform the same trade using another mechanism, whereas small institutions may accept large skipping costs as an unavoidable aspect of their everyday trading. In a recent discussion of modern financial markets, \citet{Luu:2018collateral} argued that the advent of trading platforms (such as Hotspot FX) with relatively low barriers to entry have blurred the lines between the inter-bank market and less-traditional markets. The significant heterogeneity that we observe in skipping costs is consistent with the idea that a wide and heterogeneous population of financial institutions operate on such platforms, and sometimes experience considerably different prices for similar trades.


\subsection{Price Changes}\label{sec:dP}

We now examine price changes between successive trades. Recall from Section~\ref{sec:skippingcosts} that $p_k$ denotes the price of the $k^{\text{th}}$ trade for a given currency pair on a given trading day. For a given $k$, let $p_{k'}$ denote the price of the previous trade in the same direction as the $k^{\text{th}}$ trade (e.g., both of them are buyer-initiated). Similarly, let $b_{k'}$, $a_{k'}$, and $m_{k'}$ denote the bid-, ask-, and mid-prices, respectively, immediately before the previous trade in the same direction as the $k^{\text{th}}$ trade. The \emph{change in trade price} is
\begin{equation}\label{eq:fk}f_k = \left\{\begin{array}{ll}
	\displaystyle{p_{k}-p_{k'}\,,} & \text{ if the $k^{\text{th}}$ trade is buyer-initiated} \\
	\displaystyle{p_{k'}-p_{k}\,,} & \text{ if the $k^{\text{th}}$ trade is seller-initiated\,.}\end{array}\right.
\end{equation}
Similar to Equation~\eqref{eq:rktilde}, the \emph{normalized change in trade price} is
\begin{equation}\label{eq:fktilde}
	\tilde{f}_k = \frac{f_k}{m_k}.
\end{equation}

Our results in Section~\ref{sec:skippingcosts} revealed that skipping costs vary considerably across the trades in our sample. The existence of some trades with a normalized skipping cost of several basis points suggests that, due to their CCLs, some institutions have access to a relatively small fraction of the trading opportunities that are available on the platform. This observation raises the question of how strongly CCLs impact the price changes between successive trades. This question is important: If different institutions pay considerably different prices for the same asset at a similar time, then the trade-price series may include large fluctuations that do not reflect similar changes in an asset's fundamental value. Therefore, the price-formation process on a platform that implements CCLs may be rather different than that on a platform in which all institutions can trade with all others.

To study this issue empirically, we decompose each term in the $f_k$ series into two constituent parts. We define the \emph{change in quote price} between the $k^{\text{th}}$ trade and the previous trade in the same direction by
\begin{equation}\label{eq:gk}g_k = \left\{\begin{array}{ll}
      \displaystyle{a_{k}-a_{k'}\,,} & \text{ if the $k^{\text{th}}$ trade is a buyer-initiated trade} \\
      \displaystyle{b_{k'}-b_{k}\,,} & \text{ if the $k^{\text{th}}$ trade is a seller-initiated trade\,.}\end{array}\right.
\end{equation}
We similarly calculate the \emph{change in skipping cost}: 
\begin{equation}\label{eq:hk}h_k = \left\{\begin{array}{ll}
      \displaystyle{(p_k-a_k)-(p_{k'}-a_{k'})\,,} & \text{ if the $k^{\text{th}}$ trade is a buyer-initiated trade} \\
      \displaystyle{(b_{k'}-p_{k'})-(b_{k}-p_{k})\,,} & \text{ if the $k^{\text{th}}$ trade is a seller-initiated trade\,.}\end{array}\right.
\end{equation}

For buyer-initiated trades,
\begin{equation}
f_k = p_k-p_{k'} = \left(a_k - a_{k'}\right) + \left((p_k-a_k) - (p_{k'}-a_{k'})\right) = g_k+h_k\,. \label{eq:fgh}
\end{equation}
By a similar argument, the same identity holds for seller-initiated trades. Equation (\ref{eq:fgh}) enables us to decompose each change in trade price into the constituent change in quote price and change in skipping cost.

\begin{figure}[!htbp]
\centering
\includegraphics[width=0.8\textwidth]{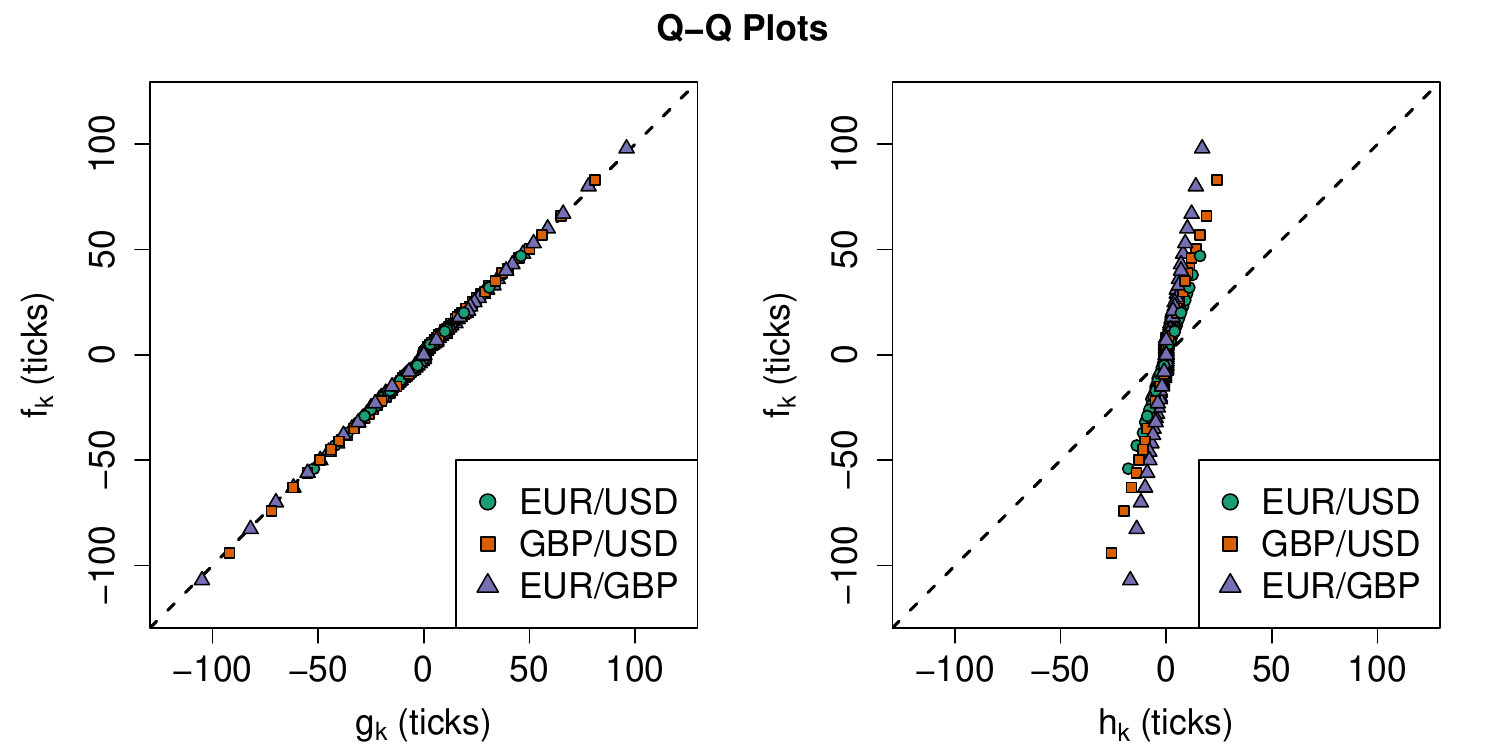}
\caption{Quantile--quantile (QQ) plots for (left) $f_k$ versus $g_k$ and (right) $f_k$ versus $h_k$. In each plot, the points indicate the percentiles of the empirical distributions. The dashed black lines indicate the diagonal.}
\label{fig:QQ}
\end{figure}

We now perform several statistical comparisons of the $f_k$, $g_k$, and $h_k$ series to quantify the relative impacts of CCLs and quote revisions on changes in trade price. In the left panel of Figure~\ref{fig:QQ}, we show a quantile--quantile (QQ) plot of the $f_k$ series versus the $g_k$ series. In this plot, the quantile points cluster tightly along the diagonal, implying that the shape of the distribution of the $f_k$ series is very similar to that of the $g_k$ series. In the right panel of Figure~\ref{fig:QQ}, we show a QQ plot of the $f_k$ series versus the $h_k$ series. In this case, the distribution of changes in skipping costs is concentrated more tightly around $0$ than is the distribution of changes in trade prices, suggesting that the changes in skipping cost account for only a small fraction of the total price change between successive trades.

\begin{figure}[!htbp]
\centering
\includegraphics[width=\textwidth]{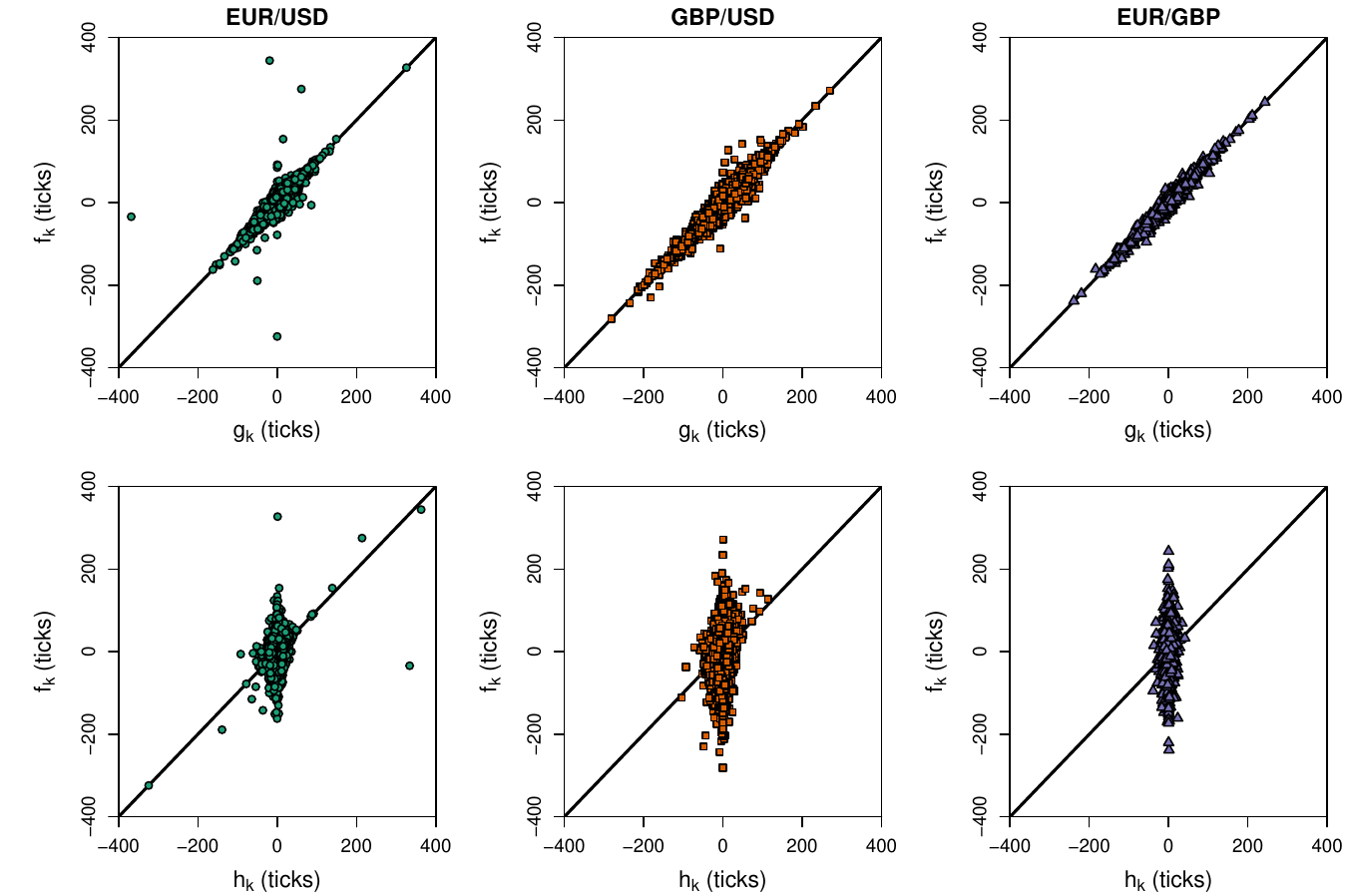}
\caption{Scatter plots of (top row) $f_k$ versus $g_k$ and (bottom row) $f_k$ versus $h_k$, for (left column) EUR/USD, (middle column) GBP/USD, and (right column) EUR/GBP. The solid black lines indicate the diagonal.}
\label{fig:PriceChangeScatters}
\end{figure}

To assess whether these results also hold at the trade-by-trade level, we make scatter plots of the individual terms of the series. In the top row of Figure~\ref{fig:PriceChangeScatters}, we show scatter plots of the $f_k$ series versus the $g_k$ series. For GBP/USD and EUR/GBP, the points cluster strongly along the diagonal, which indicates that for each trade, the change in trade price is very similar to the change in quote price. For EUR/USD, some points occur away from the diagonal, but the vast majority of points cluster along the diagonal. In the right column of Figure~\ref{fig:PriceChangeScatters}, we show scatter plots of the $f_k$ series versus the $h_k$ series. In stark contrast to the plots of $f_k$ versus $g_k$, these plots do not reveal any visible relationship between the $f_k$ and $h_k$ series for any of the three currency pairs.

To examine the relationships between the $f_k$, $g_k$, and $h_k$ series across all trades in our sample, we also calculate the sample Pearson correlation $\rho$ between these series (see Table~\ref{tab:Correlations}). When comparing the $f_k$ and $g_k$ series, we find that $\rho \approx 1$ for all three currency pairs. This quantifies the strong relationship between changes in trade price and changes in quote price (see the left panels of Figure~\ref{fig:PriceChangeScatters}) and indicates that changes in trade price are strongly correlated with corresponding changes in the underlying quotes. By contrast, when comparing the $f_k$ and $h_k$ series, we find that $\rho \approx 0$ for all three currency pairs. This provides further evidence that changes in skipping cost are uncorrelated with changes in trade price.

\begin{table}[!htbp]
\centering
\begin{tabular}{lrrr}
\hline
& EUR/USD & GBP/USD & EUR/GBP \\
\hline
\multirow{1}{*}{$f_k$ versus $g_k$} & $0.94$ \ $(<0.01)$ & $0.97$ \ $(<0.01)$ & $0.99$ \ $(<0.01)$ \\ 
\hline
\multirow{1}{*}{$f_k$ versus $h_k$} & $-0.04$ \ $(0.03)$ & $0.00$ \ $(<0.01)$ & $0.00$ \ $(0.02)$ \\
\hline
\end{tabular}
\caption{Sample Pearson correlation between (top row) changes in trade price ($f_k$) and changes in quote price ($g_k$) and (bottom row) changes in trade price ($f_k$) and changes in skipping cost ($h_k$). The numbers in parentheses are the sample standard deviations of the estimates across 10000 bootstrap samples of the data.}
\label{tab:Correlations}
\end{table}

Taken together, our results in this subsection suggest the following interpretation of Equation (\ref{eq:fgh}). Each change in trade price consists of two components: a change in the underlying quotes and a change in the skipping cost. The change in trade price is strongly correlated with the change in quotes, but it has little or no correlation with the change in skipping cost. Therefore, although the change in skipping cost sometimes constitutes a considerable fraction of the total change in trade price, this impact manifests as (additive) uncorrelated noise in the trade-price series.

From an economic perspective, the strong positive correlation between the $f_k$ and $g_k$ series and the absence of a significant correlation between the $f_k$ and $h_k$ series suggests that, during the course of everyday trading, fundamental revaluations in trade prices arise from corresponding changes in the best quotes. One can regard the $f_k$ series as a noisy observation of the $g_k$ series, where the uncorrelated, additive noise arises from the restriction of institutions' trading activities to their bilateral trading partners. The strength of this effect varies across institutions because of the heterogeneity of their CCLs.

\subsection{Volatility}\label{sec:volatility}

Our results in Section~\ref{sec:dP} suggest that changes in trade price have little or no correlation with changes in skipping cost. However, given that some trades have large skipping costs (see Section \ref{sec:skippingcosts}), it is possible that the volatility in the trade-price series differs significantly from the corresponding volatility of the underlying quotes. In this section, we assess the extent to which this is the case.

Recall from Section~\ref{sec:skippingcosts} that if the CCLs on a given platform allowed all institutions to access all trading opportunities, then all trades would occur at the best quotes, so $f_k = g_k$ for all $k$. In this case, any volatility estimate would produce the same result when applied to either of these series. However, because CCLs restrict institutions' access to liquidity, this is not true in general. By comparing the realized volatility of the trade-price series with the realized volatility of the corresponding quote-price series, it is possible to quantify the difference between the volatility in the prices that institutions actually pay for their trades and the underlying volatility that is observable in the platform-wide best quotes.

In contrast to studying the prices of individual trades, for which the application of CCLs always creates a non-negative additional cost, it is not clear \emph{a priori} whether the application of CCLs will cause the volatility of the trade-price series to be greater than or less than the volatility of the corresponding quote-price series. On one hand, it is possible for quote prices to remain stable while trade prices fluctuate. In this case, the volatility of the trade-price series would be greater than that of the quote-price series. On the other hand, it is possible for quote prices to fluctuate while trade prices remain stable. This could occur, for example, if the best quotes are populated by electronic trading algorithms that are run by institutions with very few bilateral CCLs. In this case, the quote prices would fluctuate often, but most market orders would skip these limit orders and match deeper into the LOB, such that the volatility of the trade-price series would be less than that of the quote-price series. The aim of this subsection is to determine which of these two possibilities occurs on Hotspot FX.

For each currency pair on each day, we estimate the sell-side trade-price volatility $v_A$, the sell-side quote-price volatility $v_a$, the buy-side trade-price volatility $v_B$, and the buy-side quote-price volatility $v_b$ by calculating the quadratic variation of each process, sampled at regularly spaced intervals in trade time (i.e., sampling every $x^{\text{th}}$ trade)\footnote{Consequently, the number of seconds between successive samplings is larger in periods when there are fewer trades. We repeated all of our calculations by sampling the same series at regularly spaced intervals in calendar time, and we obtain results that are qualitatively similar to those that we obtain with regularly spaced intervals in trade time.} and subsampled at regularly spaced offsets. Throughout this paper, we study buyer-initiated and seller-initiated trades separately in an attempt to disentangle our results about CCLs from the possible impact of bid--ask bounce (see \citet{Roll:1984simple}).

Following \citet{Liu:2015does}, we present our results for $K = 108$ regularly spaced sampling intervals each day (corresponding to a mean interval length of 5 minutes when measured in calendar time)\footnote{When estimating the volatility of a price series, sampling with an interval length of 5 minutes is often regarded as a simple way to reduce the impact of market microstructure noise \citep{Hansen:2006realized}.} and subsampled at $L = 10$ regularly spaced offsets. We also repeat all of our calculations for a variety of different interval numbers (ranging from $K = 50$ to $K = 500$) and a variety of different numbers of subsampling offsets (ranging from $L = 5$ to $L = 20$). Our results are qualitatively similar in all cases.

\begin{figure}[!htbp]
\centering
\includegraphics[width=\textwidth]{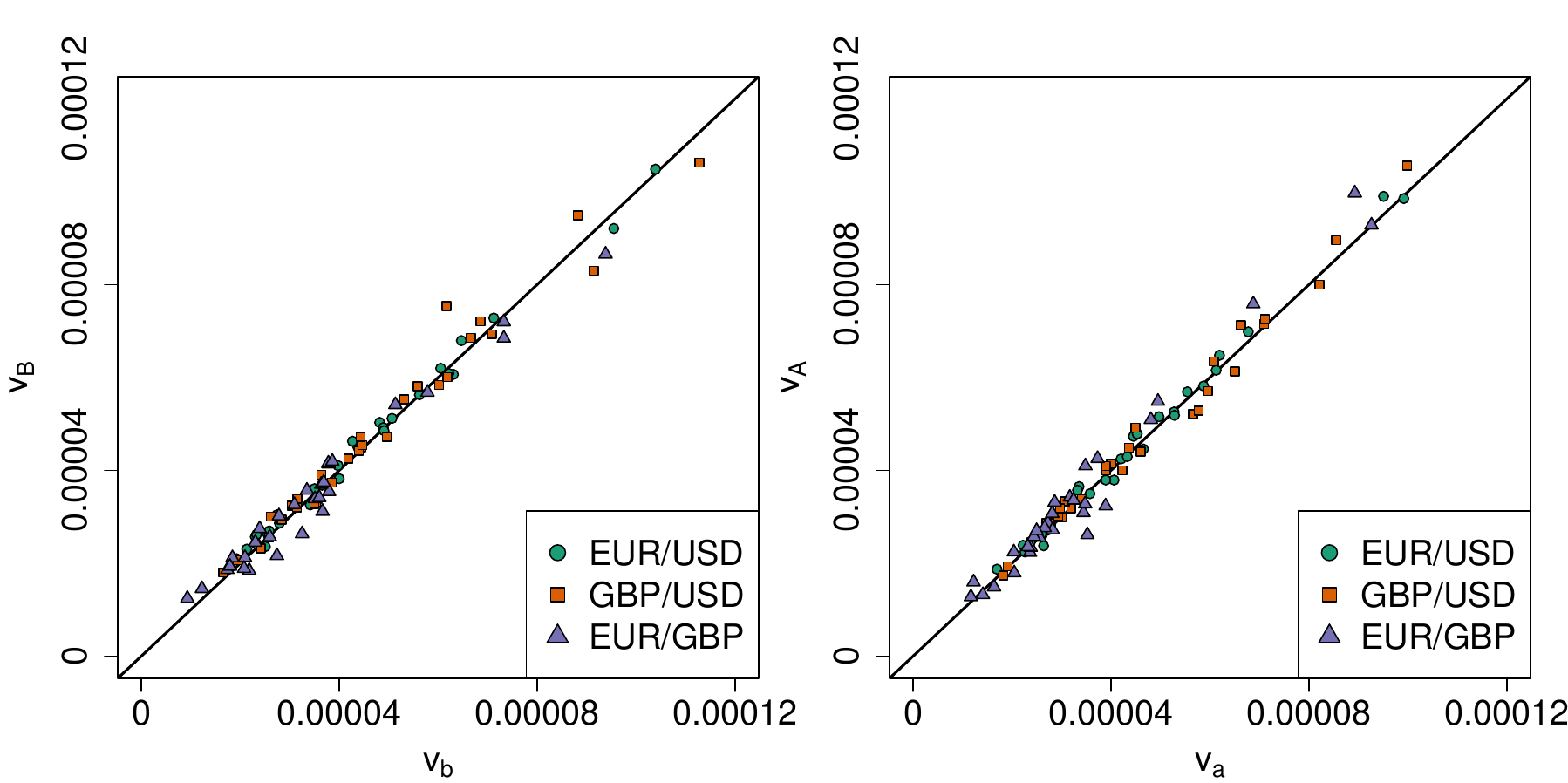}
\caption{Scatter plots of realized trade-price volatility versus realized quote-price volatility for (left panel) seller-initiated and (right panel) buyer-initiated trades. The solid black lines indicate the diagonal.}
\label{fig:VolatilityScatters}
\end{figure}

In Figure \ref{fig:VolatilityScatters}, we show scatter plots of the quote-price volatility versus the trade-price volatility for each day in our sample. The points on the scatter plots cluster along the diagonal. To help quantify the strength of this relationship, we also calculate the sample Pearson correlation $\rho$, which we measure across all 30 days in our sample (see Table \ref{tab:Correlations_Volatility}). In all cases, $\rho \approx 1$. Together, these results indicate that each day's quote-price volatility is very similar to the corresponding trade-price volatility.

\begin{table}[!htbp]
\centering
\begin{tabular}{lrrr}
\hline
& EUR/USD & GBP/USD & EUR/GBP \\
\hline
\multirow{1}{*}{$v_B$ versus $v_b$} & $0.997$ \ $(<0.01)$ & $0.985$ \ $(<0.01)$ & $0.986$ \ $(<0.01)$ \\ 
\hline
\multirow{1}{*}{$v_A$ versus $v_a$} & $0.997$ \ $(<0.01)$ & $0.993$ \ $(<0.01)$ & $0.985$ \ $(0.01)$ \\
\hline
\end{tabular}
\caption{Sample Pearson correlation between realized trade-price volatility versus realized quote-price volatility for (top row) seller-initiated and (bottom row) buyer-initiated trades. The numbers in parentheses are the sample standard deviation of the estimate across 10000 bootstrap samples of the data.}
\label{tab:Correlations_Volatility}
\end{table}

In Figure \ref{fig:VolatilityScatters}, some points lie slightly above the diagonal, and others lie slightly below the diagonal. To examine the deviation from the diagonal, we calculate the log-ratio
\begin{equation}\label{eq:z}z = \left\{\begin{array}{ll}
      \displaystyle{\log\left(v_B/v_b\right)\,,} & \text{ for seller-initiated trades} \\
      \displaystyle{\log\left(v_A/v_a\right)\,,} & \text{ for buyer-initiated trades}\end{array}\right.\end{equation}
for each currency pair on each day. A positive value of $z$ indicates that the trade-price volatility exceeds the quote-price volatility; a negative value of $z$ indicates that the quote-price volatility exceeds the trade-price volatility. For each of the three currency pairs, the mean value of $z$ is slightly positive for both buyer-initiated and seller-initiated trades (see Table~\ref{tab:mean_zs}), suggesting that, on average, the trade-price volatility on a given day typically exceeds the corresponding quote-price volatility. However, the magnitude of this effect is less than one standard deviation, so the effect is very weak. Therefore, the vast majority of volatility in the trade-price series is also directly observable in the quote-price series. This implies that the volatility observable in both the quote-price and trade-price series is dominated by a common, underlying volatility that supersedes the idiosyncratic impact of CCLs on the trade-price series.

\begin{table}[!htbp]
\centering
\begin{tabular}{lrrr}
\hline
& EUR/USD & GBP/USD & EUR/GBP \\
\hline
\multirow{1}{*}{Seller-Initiated Trades} & $0.022$ \ $(0.049)$ & $0.020$ \ $(0.064)$ & $0.010$ \ $(0.116)$ \\
\hline
\multirow{1}{*}{Buyer-Initiated Trades} & $0.015$ \ $(0.045)$ & $0.012$ \ $(0.050)$ & $0.022$ \ $(0.114)$ \\
\hline
\end{tabular}
\caption{Mean values of $z$ for (top row) seller-initiated and (bottom row) buyer-initiated trades. The numbers in parentheses are the corresponding standard deviations.}
\label{tab:mean_zs}
\end{table}

\section{A Model of Trade with CCLs}\label{sec:model}

In Section~\ref{sec:empirics}, we used data from Hotspot FX to analyze how CCLs affect everyday trading in a real market. However, studying historical data is only one aspect of examining how this mechanism may affect financial markets, and such analysis does not provide insight into how these results may change if institutions make substantial modifications to their CCLs. Because the underlying network of CCLs on Hotspot FX is fixed and unobservable to us, empirical analysis does not provide a way to investigate this important issue. Therefore, in this section, we complement our empirical work by introducing and studying a model of trade in which institutions assign CCLs to their trading counterparties.

In our model, each institution updates its buy and sell prices for a single asset and performs a trade whenever it identifies a trading counterparty that is offering to buy or sell at a mutually agreeable price. A crucial feature of the model is that not all institutions can trade with all others; instead, each institution can trade only subject to its CCLs. Therefore, trades occur at prices that depend not only on other institutions' buy and sell prices, but also on the underlying network of bilateral CCLs. 

We study a simple trading mechanism --- ignoring many features of real markets --- to highlight the relationship between the underlying CCL network (which we define formally in Section \ref{subsection:CCLnetworks}) and the trades. Our approach is similar to those of \citet{Luu:2018collateral}, \citet{Roukny:2013default}, and \citet{Weber:2017joint}, who examined how the network topology and edge density of an underlying financial network can impacts emergent properties (such as default cascades) in a simple model of trading. In our simulations (and analogous to the approaches in these papers), we fix all parameters except those that are related to the CCL network. Motivated by our empirical results in Section \ref{sec:empirics}, we assess how restricting the trading opportunities that are available to institutions affects both the prices of individual trades and the realized volatility of the quote-price and trade-price series.


\subsection{The Model}\label{sec:temporal_evolution}

The setting for our model is an infinite-horizon, continuous-time market that is populated by a set $\Theta = \left\{\theta_1, \theta_2, \ldots, \theta_N \right\}$ of $N$ institutions that trade a single risky asset. Each institution $\theta_i \in \Theta$ maintains a private buy-valuation $B_t^i$ and a private sell-valuation $A_t^i$. The values of $B_t^i$ and $A_t^i$ vary across the different institutions to reflect differences in their views of the likely future value of the asset, as well as differences in their inventory, cash-flow, financing constraints, and so on. To focus on the impact of CCLs without considering the impact of strategic activity (which could make our results more difficult to interpret), we model these prices using stylized stochastic processes. For each institution $\theta_i$, we rewrite the buy and sell prices in terms of a mid-price $M^i_t = (B_t^i+A_t^i)/2$ and spread $s^i_t = A_t^i - B_t^i$, so that
\begin{equation}\label{eq:ab}
	B^i_t = M^i_t - \frac{s^i_t}{2}\,, \quad A^i_t = M^i_t + \frac{s^i_t}{2}\,.
\end{equation}
We describe the dynamics of the spread in detail in Section \ref{sec:temporal_evolution}. For now, we remark only that we constrain the values of $s^i_t$ to never fall below the minimum value $s_0>0$.

Before simulating our model, we choose an initial state in which no trading is possible. We give details of this initialization in Section \ref{sec:parameterchoices}. Leaving aside the behavior of $s^i_t$ and $M^i_t$ at trade times, which we describe in Section \ref{sec:trading}, we assume that between trades the $s^i_t$ are governed by
\begin{equation}\label{eq:s_t}
     \mathrm{d}s^i_t = -\kappa\left(s^i_t-s_0\right)\mathrm{d}t
\end{equation}
for some constant $\kappa>0$, and we assume that the $M^i_t$ are governed by
\begin{equation}\label{eq:OUM}
     \mathrm{d}M^i_t = \gamma M^i_t\,\mathrm{d}W^{M^i}_t\,,
\end{equation}
where $\gamma>0$ is the mid-price volatility (with units of $[\text{time}]^{-\frac{1}{2}}$) and $W^{M_i}_t$ are mutually independent Brownian motions.\footnote{A possible refinement of our model is to include a common market factor $W_T$ in addition to the idiosyncratic noise terms. In that case, $\mathrm{d}M^i_t = \gamma M^i_t\left(\rho_i \,\mathrm{d}W_t + \sqrt{1-\rho_i^2} \,\mathrm{d}W^{M^i}_t\right)$, where $\rho_i$ is a Pearson correlation coefficient. We also performed simulations of this more complicated model, but we found that this additional complication adds little to our analysis. We therefore restrict our discussion to our simpler model.}

In the absence of trading, the processes $M^i_t$ are drift-free geometric Brownian motions. Equation (\ref{eq:s_t}) causes each institution's spread to revert towards its minimum value $s_0$. Our model minimizes the complications from the mixing of price and time scales in the model parameters.

Although a geometric Brownian motion with no drift has a constant mean, its variance increases with time. Without trading, our mid-prices thus spread out progressively and indefinitely over time. However, as we will see in Section \ref{sec:trading}, the occurrence of trades ensures that prices remain grouped together. By using the same values of $\gamma$ for each institution, we ensure that the only difference between different institutions is the set of trading opportunities that they can access.


\subsection{CCL Networks}\label{subsection:CCLnetworks}

We assume that each institution $\theta_i$ assigns a CCL to each other institution $\theta_j$. We assume that each institution's access to trading opportunities does not depend on time or its trading history. Therefore, for each pair of institutions, $\theta_i$ and $\theta_j$, we model the bilateral CCL with a binary indicator: either $\theta_i$ and $\theta_j$ are trading partners or they are not. For simplicity, we allow trading partners to trade arbitrarily large amounts.

In a real financial market, a pair of financial institutions can access each other's trading opportunities until they reach their bilateral CCL (see Section \ref{sec:ccls}). Therefore, our model uses a simplification of the way that CCLs operate in real markets. There are three important benefits to this simplification in our study. First, institutions in the FX spot market routinely trade huge volumes of FX each day, yet the modal size of market orders for each of the three currency pairs on Hotspot FX is just $1$ million units of the base currency. If a given pair of institutions have a sufficient bilateral CCL to access each other's trading opportunities once, then they are likely to be able to do so again. Second, in the FX spot market, trades that are agreed on day $D$ are settled on day $D+2$. Given this relatively short time interval, if an institution $\theta_i$ cannot access a trading opportunity that is offered by $\theta_j$, we claim that it is more likely that this is because their bilateral CCL is actually $0$ than because they have gradually accumulated a large exposure. Third, this simplification makes our model time-stationary. By contrast, tracking the cumulative exposure between a pair of institutions and allowing them to trade only up to their bilateral CCL yields a model that is non-stationary in time. We opt for this significant gain in model simplicity in what we regard to be an adequately realistic equilibrium-pricing framework.

We encode CCLs using an undirected,\footnote{Recall from Section \ref{sec:ccls} that the bilateral CCL between $\theta_i$ and $\theta_j$ in real markets is given by $\min\left(c_{(i,j)},c_{(j,i)}\right)$. Therefore, although it is necessary to use a directed network to model both $c_{(i,j)}$ and $c_{(j,i)}$ between individual institutions, the network of bilateral CCLs is, by definition, an undirected network.} unweighted network in which the nodes represent the institutions and the edges encode the extant bilateral credit relationships: $\theta_i$ and $\theta_j$ can trade with each other if and only if the edge $\theta_i \leftrightarrow \theta_j$ exists in the network. We call such a network a \emph{CCL network}. For any network with $N$ nodes, the maximum number of edges is $N(N-1)/2$. Therefore, a CCL network with $N$ nodes and $n$ edges has an edge density of
\begin{equation}\label{eq:d}
	d=\frac{2n}{N(N-1)}\,.
\end{equation}

\begin{figure}[!htbp]
\centering
\includegraphics[width=0.3\textwidth]{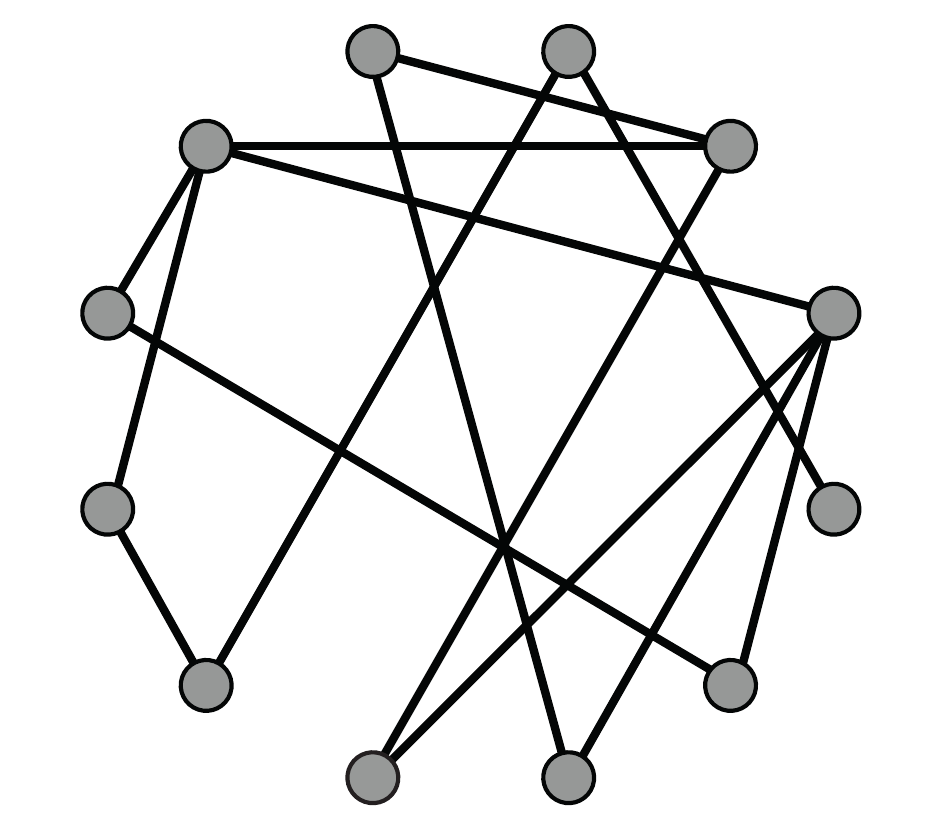}\qquad \qquad \qquad
\includegraphics[width=0.3\textwidth]{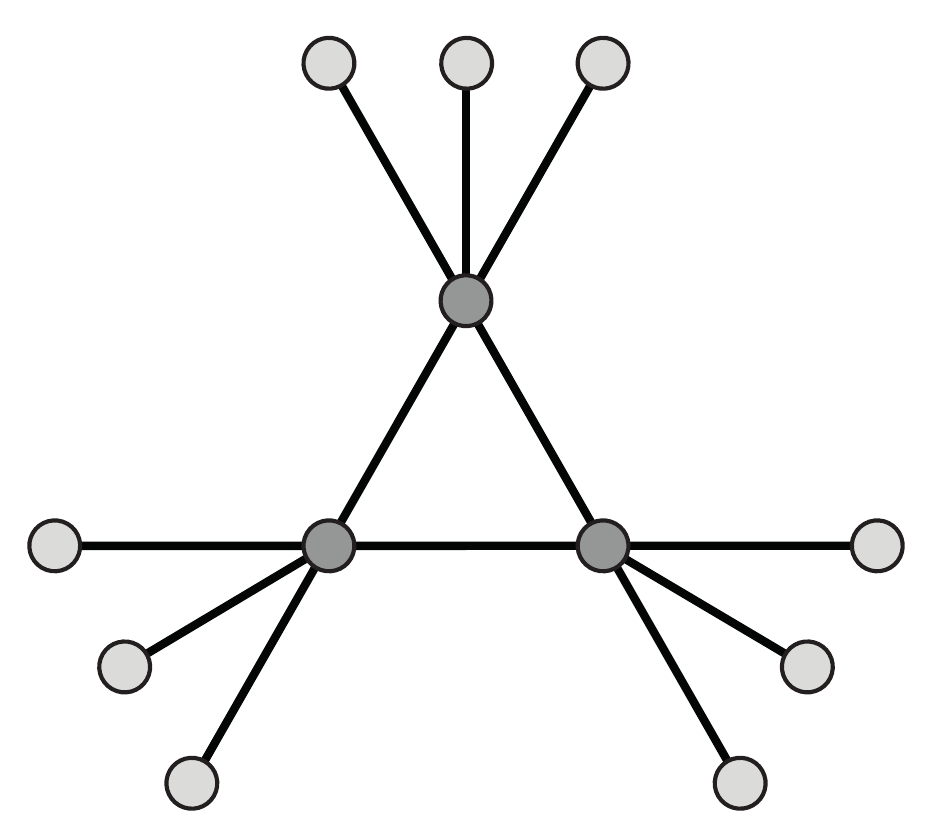}
\caption{Schematics of the two classes of CCL networks that we study. (Left) An Erd\H{o}s--R\'{e}nyi (ER) network and (right) a core--periphery (CP) network in which 3 nodes are core nodes (in dark grey) and 9 nodes are peripheral nodes (in light grey).}
\label{fig:abmtopologies}
\end{figure}

With our model, we can consider any connected CCL network with any number of nodes and any set of edges. However, to highlight the most salient features of our results, we restrict our discussion to two classes of networks with specific topological structures (see Figure~\ref{fig:abmtopologies}).

The first class of CCL networks that we consider are \emph{Erd\H{o}s--R\'{e}nyi (ER) networks} \citep{Erdos:1960evolution,newman2018}. We use this class of networks to model a market in which institutions choose their trading partners uniformly at random. Although this assumption is a poor reflection of how institutions set CCLs in a real market, studying ER networks enables us to investigate the temporal evolution of our model in a simple, stylized framework with no deterministic structure.
To construct our ER networks, we fix the edge density $d$ and use Equation (\ref{eq:d}) to calculate $n$. We place these $n$ edges uniformly at random, and we then check whether the network consists of a single connected component. If it does not, we reject the CCL network and construct another
network using the same rules. For a given value of $d$, we construct a sample of $1000$ such CCL networks, and we then simulate $1000$ independent runs of our model for each of these $1000$ CCL networks.

The second class of CCL networks that we consider are \emph{core--periphery (CP) networks} \citep{rombach2017}). Specifically, we consider CP networks that have two types of nodes: core nodes and peripheral nodes. In our CP networks, each core node is adjacent to all other core nodes, each peripheral node is adjacent to exactly one core node, and the degrees of any two core nodes differ by no more than $1$. We use these CP networks to model a market in which (1) a core group of institutions assign very high CCLs to each other, but (2) all other institutions have a credit line with just one institution (which is in the core). Several recent studies (see, e.g., \citet{Craig:2014interbank,Fricke:2012core,Iori:2008network}) have suggested that many large financial markets have an approximate CP structure, with a core that consists of large banks and a periphery that consists of smaller financial institutions, such as small banks, hedge funds, or mutual funds. Our CP structure is a stylized (and deterministic) version of such a structure.

To construct our CP networks, we first fix the fraction $\psi$ of peripheral nodes. When $\psi=0$, all institutions are core institutions, so the CCL network is complete and all institutions are able to trade with all others. For a given choice of $\psi$ (and therefore of $d$), we construct a single CCL network, and we then simulate $1000$ independent runs of the model for this CCL network.

\subsection{Trading}\label{sec:trading}

We assume that a trade occurs at each time $t^*$ such that a pair of institutions, $\theta_i$ and $\theta_j$, with $\theta_i \leftrightarrow \theta_j$, have prices that satisfy $B^i_{t^*}=A^j_{t^*}$. We call this price the \emph{trade price}. To aid comparisons between the output of our model and our empirical results in Section \ref{sec:empirics}, we use the following rule to classify trades as buyer-initiated or seller-initiated. Let $\bar{M}_t$ denote the mean\footnote{For simplicity, we use an unweighted average. It is possible to extend our model by instead using a weighted average to reflect possible asymmetries between different institutions. For example, one can assign a larger weight to larger institutions.} of the $N$ institutions' mid-prices at time $t$. Consider a trade that occurs between $\theta_i$ and $\theta_j$ at time $t^*$ with trade price of $p=B^i_{t^*}=A^j_{t^*}$. If $p \geq \bar{M}_t$, we label this trade as buyer-initiated, and we call $\theta_i$ the \emph{initiator} and $\theta_j$ the \emph{acceptor}. Otherwise, we label this trade as seller-initiated, and we call $\theta_j$ the initiator and $\theta_i$ the acceptor.

For each trade, we think of the initiator as having submitted a market order at the trade price and we think of the acceptor as having owned a limit order --- which is then matched by this market order --- at the trade price. The fewer bilateral CCLs that the initiator has, the more that
we expect this price to be from $\bar{M}_t$. We thereby encode the relative competitive disadvantage of institutions with poor bilateral CCL connections.

Whenever a trade occurs between a pair of institutions, $\theta_i$ and $\theta_j$, we record the skipping cost of the trade and then move $M^i_{t^*}$ and $M^j_{t^*}$ together by $s_0$ to reflect that $\theta_i$ and $\theta_j$ have successfully satisfied their desires to trade with each other. If the trade is buyer-initiated, we subtract $s_0/2$ from $M^i_{t^*}$ and add $s_0/2$ to $M^j_{t^*}$; if the trade is seller-initiated, we instead subtract $s_0/2$ from $M^j_{t^*}$ and add $s_0/2$ to $M^i_{t^*}$. We then increase both $s^i_t$ and $s^j_t$ by $s_0/2$. This adjustment models a decrease in trading desire from the initiator and acceptor from the execution of the trade (as reflected by widening the bid--ask spread), and it removes the undesirable possibility of the initiator's price and acceptor's price being equal infinitely often in an arbitrarily small time interval. When we have completed these updates, we recalculate the values of $B^i_{t^*}$, $A^i_{t^*}$, $B^j_{t^*}$, and $A^j_{t^*}$ using Equation \eqref{eq:ab}.

We now see how trading stops the mid-prices from spreading out indefinitely over time. If the distance between the mid-prices of $\theta_i$ and $\theta_j$ equals the mean of their spreads, then the buy price of one trader meets the sell price of another and a trade occurs.


\subsection{Implementation and Parameter Choices}\label{sec:parameterchoices}

We simulate the evolution of our model in discrete time, with a time step $\Delta t >0$, using a simple explicit (Euler--Maruyama) difference scheme. This discretization produces an overshoot before we detect that a trade should take place. Therefore, whenever a buyer-initiated trade occurs between a buyer $\theta_i$ and a seller $\theta_j$, we actually observe $B^i_t>A^j_t$, rather than $B^i_t=A^j_t$. In the simplest (and, for small spreads, generic) case, no other relevant prices are sandwiched between these buy and sell prices. Whenever this happens, we deem a trade to have taken place at the end of the time step and at a price $p= \left(B^i_t+A^j_t\right)/2$. In a very small number of cases, the discrete time-stepping may create more than one trading opportunity in a single time step. In such a case, we first deal with the trade that occurs furthest from $\bar{M}_t$. After recording this trade and updating the buyer's and seller's prices (see Section \ref{sec:trading}), we then process the trading opportunity whose trade price is furthest from the updated $\bar{M}_t$, then the opportunity with the second-furthest trading price, and so on until there are no further trading opportunities.\footnote{It is possible to avoid this situation by implementing an algorithm to search for the exact time $t$ such that $B_t^i = A_t^j$. However, as we discuss later in this subsection, we choose a sufficiently small value of $\Delta t$ to ensure that it is rare for multiple trades to occur within a given time step. We therefore regard our choice of using fixed-length time steps to be a sensible heuristic.}

Because we aim to investigate how CCLs affect skipping costs, we fix the values of $\gamma$, $\kappa$, and $s_0$ and study our model's output (and how it varies) for different CCL networks (see Section \ref{subsection:CCLnetworks}). We set $s_0 = \epsilon M_0$ with $\epsilon = 0.001$, which implies spreads of about $0.1\%$. We choose $\kappa = 1$, which sets the (otherwise arbitrary) time unit as $1/\kappa=1$. We set $\gamma = \epsilon\sqrt{\kappa} = 0.001$ to balance the changes in the spread and the changes in the mid-price.

We initialize the mid-prices $M^i$ by drawing them randomly from a normal distribution with mean $1$ and standard deviation $\epsilon$. We then run the trade-processing algorithm that we described in Section \ref{sec:trading} (but without actually recording any trades) to adjust the mid-prices and spreads of all institutions for whom this initial state would cause trading to occur. We repeat this step until no trading opportunities remain (i.e., until $B^i_{t_0}<A^j_{t_0}$ for each pair of institutions, $\theta_i$ and $\theta_j$, for which $\theta_i \leftrightarrow \theta_j$).

The final parameter in our model is the discrete time step $\Delta t$. The dominant term in the discrete temporal evolution of the system is the noise term, which in relative terms (i.e., relative to the value of the relevant quantity at the beginning of the time step) is $O(\gamma\sqrt{\Delta t})$. Accurate discretization of the stochastic processes requires this term to be small. Moreover, we wish to avoid the situation in which the discrete time steps regularly create multiple simultaneous trading opportunities. We expect the separation of the mid-prices to be $O(\epsilon M_0/N)$. We would like this separation to be several times the standard deviation of the noise term, such that the probability of observing a discrete price change that is larger than this is very small. We therefore take $\Delta t = 1/(3N^2)$. This choice of $\Delta t$ is also sufficiently small for us to be able to neglect errors that are associated with the numerical integration of the stochastic differential equations \eqref{eq:OUM}.

\begin{figure}[!htbp]
\centering
\includegraphics[width=0.9\textwidth]{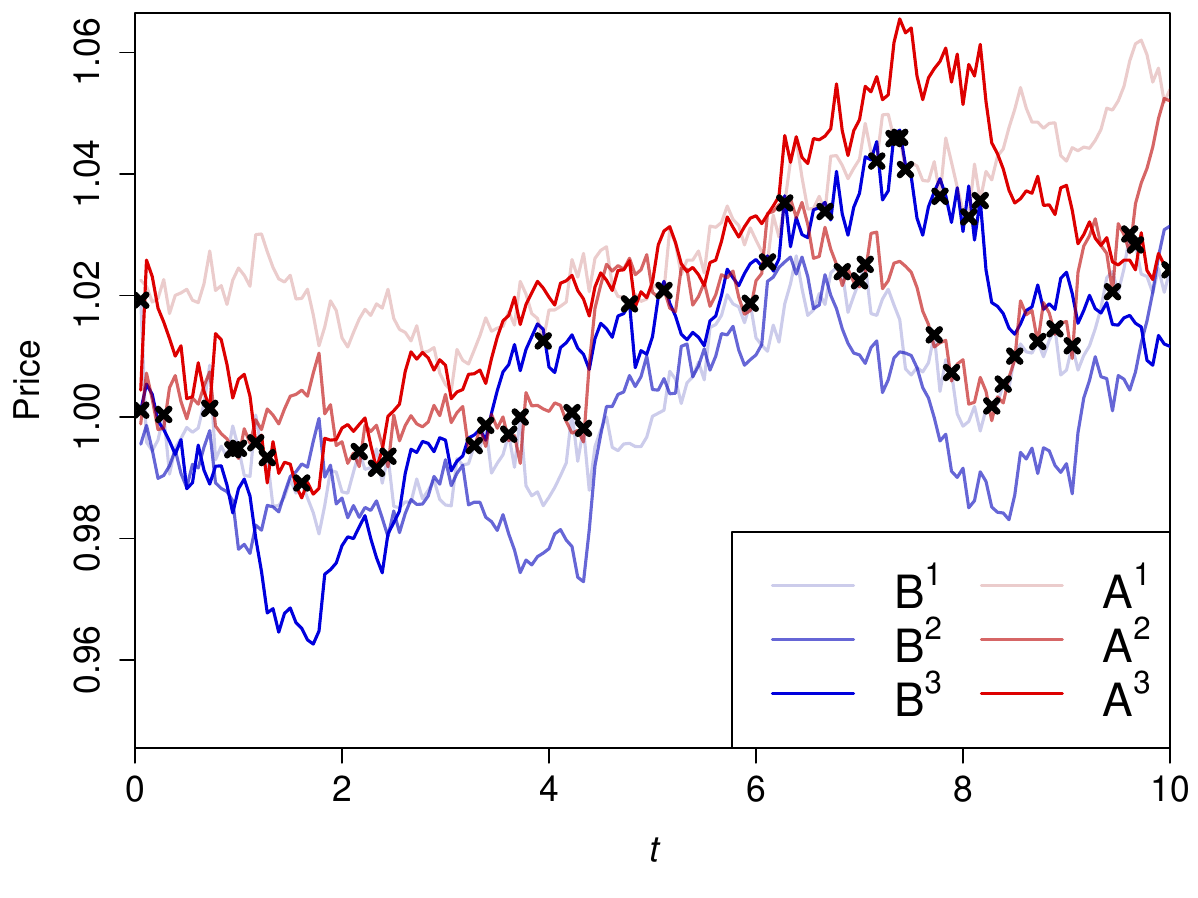}
\caption{An example simulation of our model using a CCL network with $N=3$ institutions, the parameter values in Section~\ref{sec:parameterchoices}, and a CCL network in which $\theta_1 \leftrightarrow \theta_2$ and $\theta_1 \leftrightarrow \theta_3$, but in which $\theta_2$ and $\theta_3$ cannot trade with each other. The black crosses indicate trades.}
\label{fig:ExampleSimulation}
\end{figure}

For the simulation results that we present in Section~\ref{sec:modelresults}, we consider networks with $N=128$ institutions.\footnote{We also conducted simulations with several different choices of $N$ between 100 to 1000 (with appropriately modified values of $\Delta t$). Our results are qualitatively similar in each case.} For each CCL network that we study, we simulate the temporal evolution of our model from $t=0$ to $t=10$. We discard all activity before $t=2$ as a burn-in period to avoid incorporating the transient behavior. We verified that these choices are sufficiently large by examining results using a variety of different burn-in periods and total time lengths. Our results are similar for all burn-in periods that are longer than about $t=1$ 
and for all total time periods that are larger than about $t=2$. In Figure~\ref{fig:ExampleSimulation}, we show a single simulation of our model to illustrate how heterogeneity in institutions' access to trading opportunities (which arise as a direct consequence of their CCLs) manifest in the trade-price series.

\subsection{Simulation Results}\label{sec:modelresults}

We study buyer-initiated and seller-initiated trades separately via the trade-classification algorithm that we described in Section \ref{sec:trading}. In line with our expectations (due to the symmetry of buyers and sellers in our model), our results are qualitatively the same for buyer-initiated and seller-initiated trades. To increase the size of our samples, we aggregate buyer-initiated and seller-initiated trades and present our results for all trades together.

\begin{figure}[!htbp]
\centering
\includegraphics[width=0.45\textwidth]{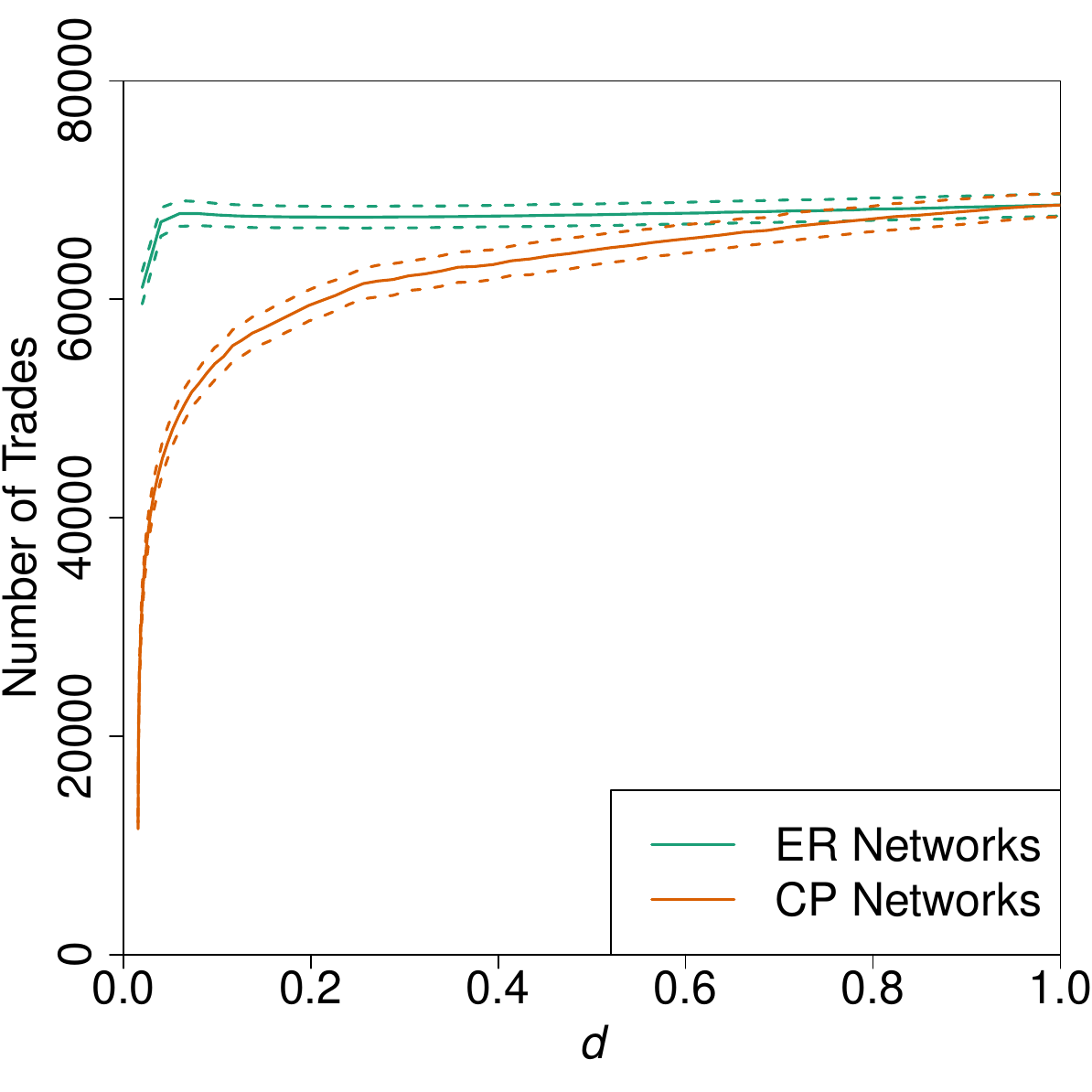}
\includegraphics[width=0.45\textwidth]{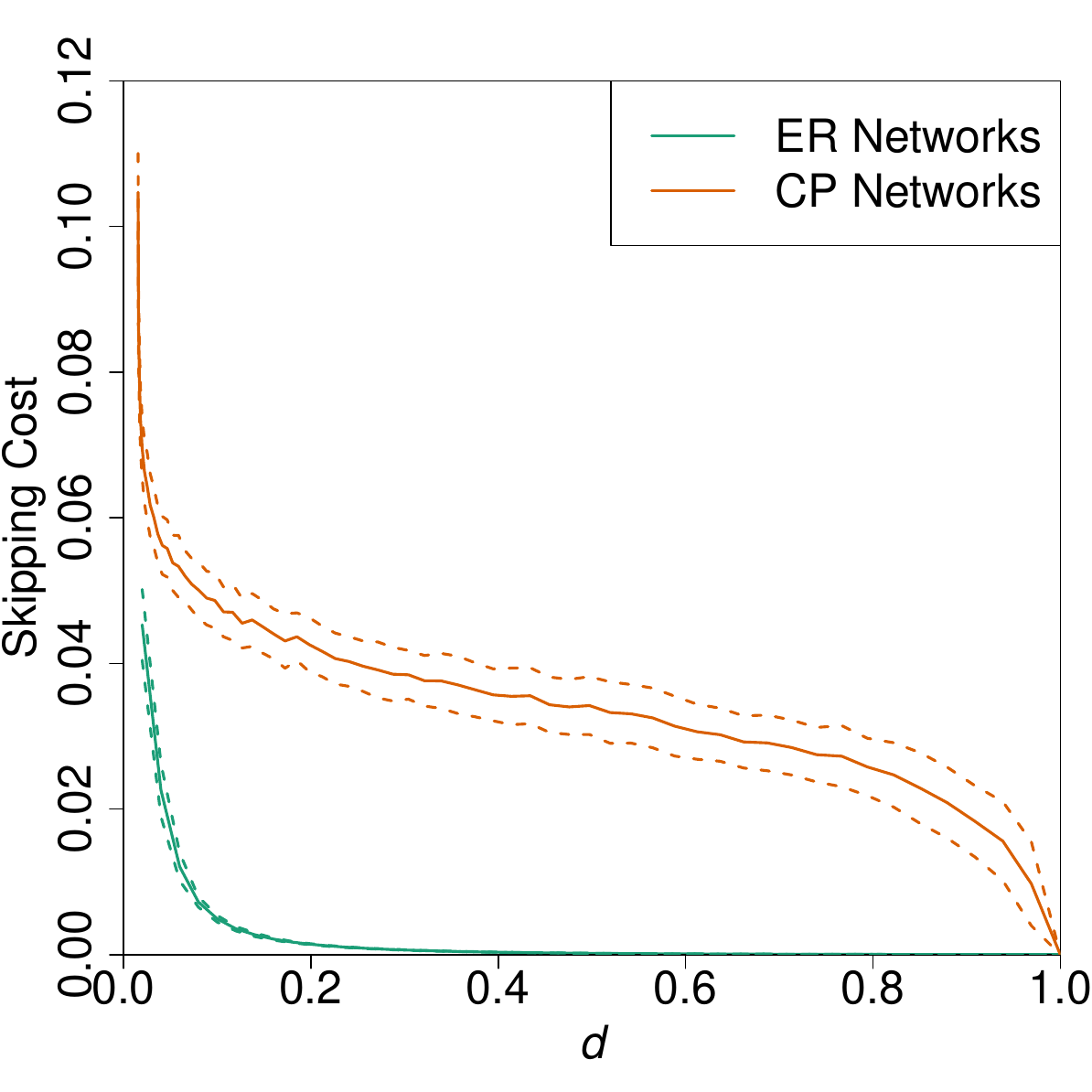}
\caption{(Left) Number of trades and (right) skipping costs for (green) ER and (orange) CP networks. The solid curves indicate the mean across all independent runs of our model. The dashed curves indicate one standard deviation from the mean.}
\label{fig:TradeCountSimulation}
\end{figure}

We first study the number of trades and the skipping costs for each edge density $d$ (see Figure~\ref{fig:TradeCountSimulation}). For both ER and CP networks, CCL networks with progressively lower edge densities result in progressively fewer trades and progressively larger skipping costs. The intuition is simple: The lower the edge density, the smaller the number of bilateral trading partners in a population. This, in turn, leads to a smaller number of trades and larger skipping costs. More interestingly, the mean number of trades is greater and the mean skipping cost is smaller for an ER network than for a CP network with the same edge density. From a practical perspective, this suggests that the influence of CCLs depends not only on the number of trading partners of each institution, but also on the network topology.

For both classes of networks, the mean skipping cost decreases rapidly as $d$ increases from $0$ to about $0.1$. For ER networks, the mean skipping cost is very close to $0$ for all values of $d$ above about $0.3$. In this regime, CCLs have a very small impact on individual trade prices. For CP networks, the mean skipping cost is very large for very small values of $d$, but it decreases to $0$ as $d$ approaches $1$ (for which the CCL network is complete, so all trades have $0$ skipping cost by definition). Moreover, the standard deviation of skipping costs is much larger for CP networks than it is for the ER networks.

\begin{figure}[!htbp]
\centering
\includegraphics[width=0.45\textwidth]{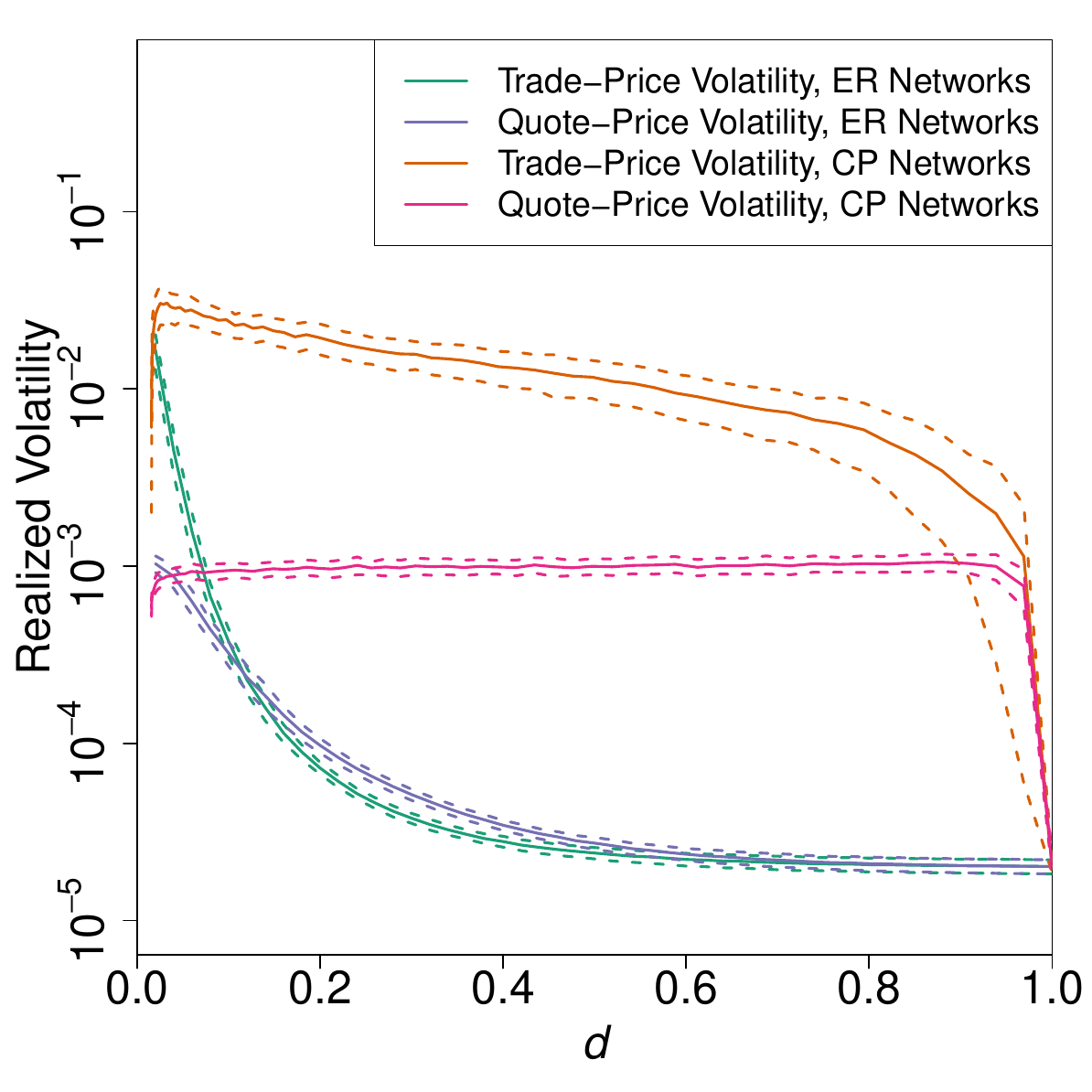}
\includegraphics[width=0.45\textwidth]{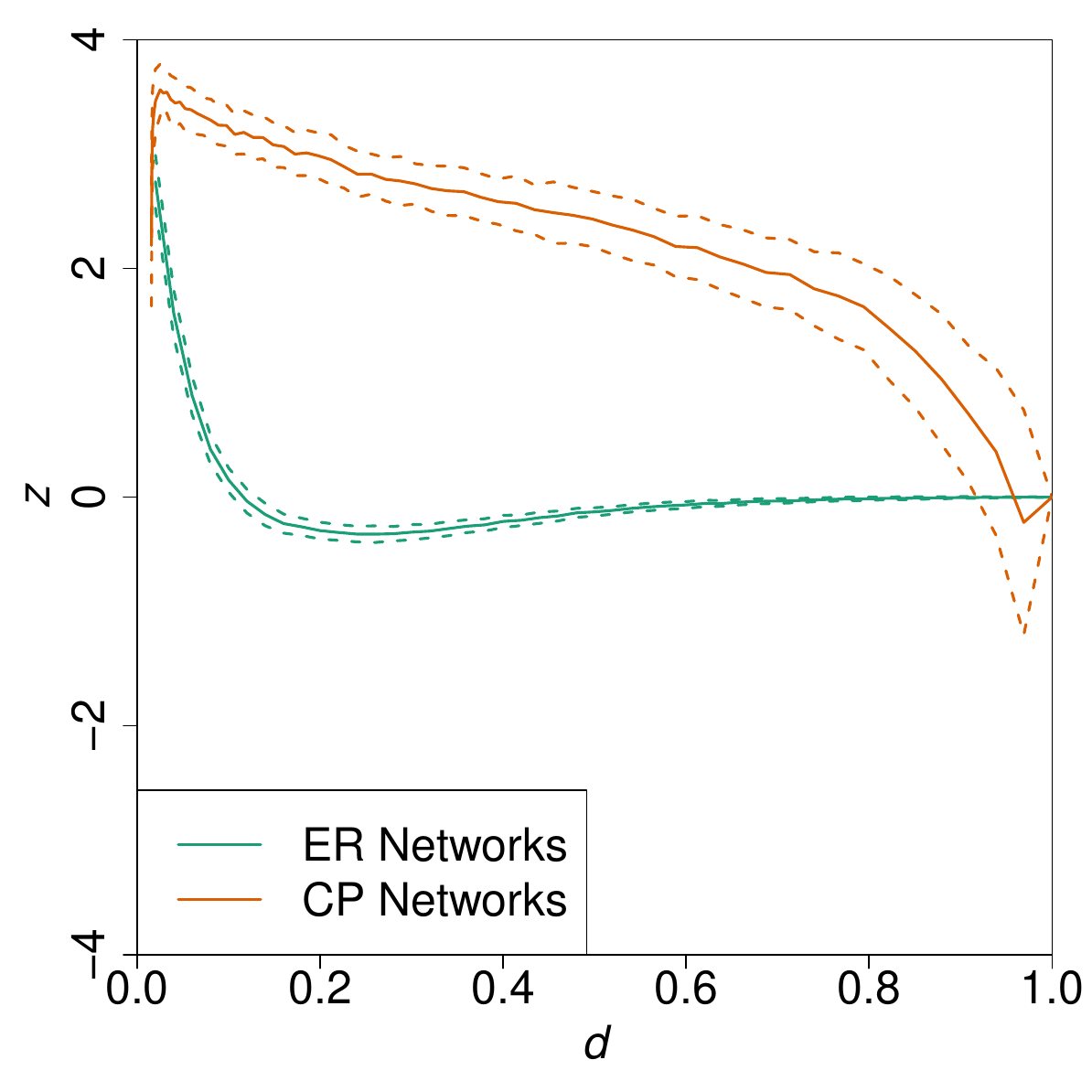}
\caption{(Left) Realized trade-price and quote-price volatilities and (right) mean of the log-ratio $z$ (see Equation \eqref{eq:z}). The solid curves indicate the mean across all independent runs of our model. The dashed curves indicate one standard deviation from the mean.}
\label{fig:VolatilitySimulation}
\end{figure}

In Figure~\ref{fig:VolatilitySimulation}, we plot the trade-price and quote-price volatilities for each $d$. We describe our methodology for measuring realized volatility in Appendix~B.\footnote{When studying our model, we only consider realized volatility computed based on event-time sampling. We did not repeat our calculations using calendar-time sampling, because our choice of time scale is arbitrary.}

For the ER networks, the trade-price volatility exceeds the quote-price volatility when $d \lessapprox 0.1$. As $d$ increases, the trade-price volatility decreases faster than the quote-price volatility. This observation is intuitively sensible because quote-price volatility is determined by the most extreme prices (i.e., the maximum among all buy prices and the minimum among all sell prices). An increase in $d$ creates relatively few additional edges for each institution, including for the institutions whose prices reside at these extremes. Therefore, it has a relatively modest impact on quote-price volatility. By contrast, trade-price volatility depends on the prices of all trades that are conducted by all institutions, which in turn are affected by all bilateral CCLs between all pairs of institutions. Consequently, trade-price volatility is influenced much more strongly than quote-price volatility by $d$.

For the CP networks, quote-price volatility is approximately constant across all values of $d$ except for those that are close to $0$ or $1$. By contrast, trade-price volatility first increases sharply as $d$ increases slightly above $0$, and it subsequently decreases gradually and then decreases sharply as $d$ increases beyond about $0.95$. This result illustrates that for the CP networks, edge density has a much stronger influence on trade-price volatility than it does on quote-price volatility.

Together, our simulations suggest that CCLs can significantly impact both trade prices (see Figure~\ref{fig:TradeCountSimulation}) and volatility (see Figure~\ref{fig:VolatilitySimulation}). Two features are particularly interesting. First, as the edge density of a CCL network decreases, both the skipping costs of individual trades and the trade-price volatility increase. This increase is not accompanied by a similar increase in quote-price volatility. Second, the impact of CCLs depends not only on edge density, but also on the specific topology of a CCL network. Therefore, forecasting how a difference in the edge density $d$ impacts skipping costs and trade-price volatility also requires knowledge of a CCL network's topology. Intuitively, this result implies that understanding the possible impact of CCLs in financial markets requires knowledge not only of how many institutions are trading partners, but also of which institutions are trading partners with each other.


\section{Conclusions and Discussion}\label{sec:conclusion}

We investigated how the application of CCLs impacts the prices that institutions pay for their trades during everyday trading. We first examined this issue empirically by studying a data set from Hotspot FX, which is a large electronic trading platform in the FX spot market that utilizes CCLs. Although we observed that CCLs have little or zero impact on most of the trades in our sample, we also identified a handful of trades for which the application of CCLs created very large skipping costs. We argued that this is a natural consequence of the heterogeneity in the types and sizes of institutions that trade on Hotspot FX. By implementing CCLs, Hotspot FX can facilitate trade for a wide variety of different financial institutions while providing these institutions with the ability to decide for themselves whether to trade with specific counterparties. Because of this direct control of counterparty exposures, there is no need for Hotspot FX to set high barriers to entry for new participants. Indeed, two Triennial Central Bank Surveys from Bank for International Settlements have noted that a new trend for direct participation from small, non-bank institutions has been a key driver for sustained growth in FX volumes \citep{BIS:2010triennial,BIS:2013triennial}. Our findings are consistent with the hypothesis that a wide variety of different financial institutions, with access to different sets of trading opportunities, interact simultaneously on Hotspot FX.

We also considered how CCLs impact volatility. By decomposing price changes into two components (one that is attributable to changes in the best quotes and another that is attributable to contemporaneous changes in skipping costs), we found empirically that CCLs contribute a small additional volatility to the trade-price series on Hotspot FX.

To complement our empirical analysis, we also introduced a model of a single-asset market in which institutions assign CCLs to their trading counterparties. In our model, the network of CCLs gives explicit control over the interaction topology between different institutions. By fixing the model parameters and varying only the CCL network, we studied how CCLs impact trade in our artificial market. Our main observation is that both edge density and the network topology are important for determining the skipping costs of trades and the corresponding volatility in the trade-price series. This presents a difficulty for regulators: How can one monitor the state of the CCL network between institutions in real time? In our view, to paint a realistic picture of market dynamics, policies that seek to improve financial stability must consider this important question.

To the best of our knowledge, our paper is the only empirical or theoretical study to explore the use of the CCL mechanism in a quantitative framework. Our results help to illuminate several important questions about the impact of CCLs on everyday trading. Our empirical results indicate that CCLs do not strongly impact the prices of the vast majority of trades during everyday trading on Hotspot FX. We therefore argue that one can regard the application of CCLs (and the consequent creation of skipping costs) as a reasonable cost of providing direct market access to a broad selection of different financial institutions, rather than as a weakness of this market design. However, the simulations of our model suggest that skipping costs and trade-price volatility can escalate rapidly as CCLs become more restrictive. Therefore, much like credit-valuation adjustments and trade novation via a central counterparty (see Section~\ref{sec:lit_rev}), CCLs do not provide a simple solution to the problem of counterparty credit risk.

It is interesting to compare our empirical results (see Section~\ref{sec:empirics}) to our simulation results (see Section~\ref{sec:model}). As a thought experiment, if we were to assume that our model is the exact data-generating mechanism for the trades that we observe empirically and that the only possible classes of CCL networks are ER networks and CP networks, then we could use our empirical results to draw inferences about the underlying CCL network on Hotspot FX. In this highly stylized framework, the small mean skipping cost of trades (see Section~\ref{sec:skippingcosts}) and the small mean log-ratio of trade-price to quote-price volatility (see Section~\ref{sec:volatility}) are both consistent with the conclusion that the CCL network on Hotspot FX is an ER network whose edge density is larger than about $0.1$. (See Figures~\ref{fig:TradeCountSimulation} and \ref{fig:VolatilitySimulation}.) Reflecting on this thought experiment yields two contrasting points. On one hand, it is clear that neither of the thought experiment's assumptions hold in reality, because trades on Hotspot FX occur via a process that is much more complicated than the one in our model and real CCL networks are not constrained to be ER or CP networks. On the other hand, the qualitative message from our thought experiment is intuitively reasonable: If the CCL network on Hotspot FX has a very small edge density, then we would likely observe many more trades with very large skipping costs. Therefore, the true CCL network is unlikely to be extremely sparse. Similarly, if the CCL network obeys a deterministic structure, then we would likely observe some basic hallmarks when performing a statistical analysis of individual trades. For example, if the CCL network has a perfect partitioning into core institutions and peripheral institutions, there likely would be a clustering of trades into core--core trades and core--periphery trades. Given that no such clustering is apparent anywhere in our empirical results (see Figures~\ref{fig:Skipping}, \ref{fig:QQ}, and \ref{fig:PriceChangeScatters}), we conclude that an ER network may be a better choice than a CP network (of the form that we studied) as a stylized model of a CCL network.

It is also important to compare our results to those of other studies of how financial networks can impact trade. As we discussed in Section~\ref{sec:lit_rev}, \citet{Luu:2018collateral} investigated the dynamics of collateral in the presence of rehypothecation and reported that contagion effects vary much more rapidly as a function of edge density for CP networks than they do for ER networks. Consistent with these findings, we also find that the impact of CCLs on trade prices varies much more sharply as a function of edge density for CP networks than it does for ER networks. \citet{Luu:2018collateral} argued that ``network structures with highly concentrated collateral flows [such as CP networks] are \ldots \,characterised by a trade-off between liquidity and systemic risk''. In other words, CP networks are preferable for increased liquidity, but ER networks are preferable for less spreading of a default contagion. Our results illustrate that, on average, a CP network with a given edge density produces a larger impact on trade prices (via larger skipping costs) than an ER network with the same edge density.

It is also interesting to compare our findings to those of \citet{Roukny:2013default}, who studied how a financial network's topology can impact cascades of defaults. Roukny et al. reported that network topology does not heavily influence default cascades when considering only load-redistribution (i.e., diversification) effects, but that it can be an important factor in the presence of a contagion. In our model, a larger edge density leads to a lower mean skipping cost (much like the version of the \citet{Roukny:2013default} model that considers only load-redistribution effects). Our simulation results also illustrate that the topology of a CCL network can significantly impact both skipping costs (see Figure~\ref{fig:TradeCountSimulation}) and the corresponding volatility of the trade-price and quote-price series (see Figure~\ref{fig:VolatilitySimulation}). \citet{Roukny:2013default} argued that ``hub'' nodes (i.e., nodes that have a large degree) contribute both to improving market stability (by diversifying shocks) and to impairing it (by amplifying contagions of defaults). In our model, hub nodes exist in the core of a CP network with low edge density. Our simulations reveal that these types of CCL networks can lead to very large skipping costs. Intuitively, this makes sense: In the context of our model, hubs provide trading opportunities to a wide range of other financial institutions, many of which have relatively poor access to other institutions' trading opportunities. This finding is consistent with skipping costs being a reasonable cost of providing direct market access to a broad selection of different financial institutions.

There are many possible extensions to our model. For example, one can modify the temporal evolution of institutions' buy and sell prices to more closely reflect behavior in real markets. Possible extensions in this direction include incorporating stochastic volatility and exogenous jump discontinuities that affect all institutions. As another example, one can relax the assumption that different institutions' $M_t^i$ series are independent, such as by incorporating a common dependence on the prices of recent trades (which are visible to all participants on Hotspot FX via the trade-data stream, as we discussed in Section \ref{sec:ccls}). As a third example, one can extend our model to incorporate multiple assets, such as several different currency pairs. In the context of the FX spot market, this extension may provide insight into the possible emergence of arbitrage opportunities and more generally into how the efficiency of financial markets can emerge endogenously from interactions between many heterogeneous institutions. As a fourth example, one can incorporate strategic considerations into our model. For instance, different institutions can implement different time-update rules for their buy and sell prices to reflect heterogeneity in their trading styles to reflect the buy or sell prices of other institutions with which they possess a CCL (and whose $B_t^i$ and $A_t^i$ series they can therefore observe in real time) or to reflect their estimated probabilities of experiencing a counterparty default. The implementation of CCLs can thereby create an interesting feedback loop in each institutions' buy and sell prices: By using CCLs to prevent trading with counterparties that a financial institution perceives to be unreasonably likely to default, it may choose to offer a smaller spread than would otherwise be the case. In short, in our model, the progressively restrictive application of CCLs progressively increases the skipping costs of trades; however, in real markets, the ability to implement CCLs may actually convince some institutions to narrow their spread and thereby offer other institutions better prices than they would do otherwise.

We have examined the question of how CCLs impact trade prices during everyday trading. An important topic for future work is to analyze CCLs during periods of market stress. Specifically, it is important to assess whether institutions modify their CCLs during turbulent periods to reflect the heightened probability of experiencing a counterparty failure. It is also important to examine whether (and when) such modifications significantly impact the statistical properties of the trade-price series.

Another important open question is how one might implement CCLs alongside other measures to mitigate counterparty credit risk. For example, a platform can offer institutions the ability to apply CCLs \emph{and} novate trades via a CCP. However, before such a configuration can be adopted, there is an important question to address: How should trades that are novated by the CCP count towards an institution's CCLs? One possible solution is that institutions can have a CCL with the CCP itself to guard against the possibility that the CCP fails. Given the relatively low impact of CCLs that we observed on Hotspot FX, we encourage further research in this area to help improve understanding of this interesting but hitherto unexplored market mechanism.


\section*{Acknowledgements}

We thank Franklin Allen, Bruno Biais, Julius Bonart, Jean-Philippe Bouchaud, Yann Braouezec, Damiano Brigo, Rama Cont, Jonathan Donier, J. Doyne Farmer, Ben Hambly, Charles-Albert Lehalle, Albert Menkveld, Stephen Roberts, Cosma Shalizi, Thaleia Zariphopoulou, and Ilija Zovko for helpful comments. MDG gratefully acknowledges support from the James S. McDonnell Foundation, the Oxford--Man Institute of Quantitative Finance, and the EPSRC (through Industrial CASE Award 08001834).

\appendix

\section*{Appendix A: The Hotspot FX Data}\label{app:data_analysis}

In this appendix, we describe the data that we examine in our empirical analysis. We give more details, including an explicit description of our data-processing algorithms, in \citet{Gould:2016quasi}.

During our sample period, three major multi-institution trading platforms dominated electronic trading volumes in the FX spot market: Reuters, Electronic Broking Services (EBS), and Hotspot FX.\footnote{See \citet{Bech:2012fx} for an estimated breakdown of transaction volumes between platforms during this period. Since then, the market share of Hotspot FX (which is now called ``Cboe FX Markets'') has increased considerably.} 
All three of these platforms use similar trading mechanics; in particular, all three implement CCLs via QCLOBs. Importantly, however, EBS and Reuters primarily serve the interbank market, whereas Hotspot FX serves both the interbank market and a broad range of other financial institutions, such as hedge funds, commodity trading advisers, corporate treasuries, and institutional asset managers.

Hotspot FX operates continuous trading for 24 hours each day and 7 days each week. However, the vast majority of activity on the platform occurs during the {peak trading hours} of $08$:$00$:$00$--$17$:$00$:$00$ GMT from Monday to Friday. We exclude all data from outside these time windows to ensure that our results are not influenced by unusual behavior during inactive periods. We exclude 3 May (i.e., the May Bank Holiday in the UK) and 31 May (i.e., the Spring Bank Holiday in the UK and Memorial Day in the US) because the market activity on these days was extremely low. We also exclude the 11 days that include a gap in recording that lasts 30 seconds or more. After making these exclusions, our data set contains the peak trading hours for each of 30 trading days. In Table \ref{tab:TradeSummaryStats}, we give summary statistics for each of the three currency pairs. Consistent with the market-wide volume ratios that were reported by the \citet{BIS:2010triennial}, the mean daily volume of market orders for EUR/USD exceeds that for GBP/USD by a factor of about $3$ and that for EUR/GBP by a factor of about $9$.

\begin{table}[!htbp]
\small
\centering
\begin{tabular}{|l|rrr|rrr|}
  \hline
  & EUR/USD & GBP/USD & EUR/GBP & EUR/USD & GBP/USD & EUR/GBP \\
  \hline
  & \multicolumn{3}{c|}{Panel A: Volume of Market Orders} & \multicolumn{3}{c|}{Panel B: Volume of Limit Orders} \\
  \hline
  Min & $2.5\times 10^{9}$ & $7.4\times 10^{8}$ & $1.0\times 10^{8}$ & $7.2\times 10^{12}$ & $5.5\times 10^{12}$ & $3.7\times 10^{12}$ \\ 
  Median & $4.4\times 10^{8}$ & $1.5\times 10^{9}$ & $3.6\times 10^{8}$ & $9.4\times 10^{12}$ & $7.8\times 10^{12}$ & $6.2\times 10^{12}$ \\ 
  Max & $7.5\times 10^{9}$ & $2.5\times 10^{9}$ & $1.2\times 10^{9}$ & $1.4\times 10^{13}$ & $9.7\times 10^{12}$ & $7.6\times 10^{12}$ \\ 
  Mean & $4.6\times 10^{9}$ & $1.5\times 10^{9}$ & $4.0\times 10^{8}$ & $1.0\times 10^{13}$ & $7.9\times 10^{12}$ & $6.2\times 10^{12}$ \\ 
  St. Dev. & $1.2\times 10^{9}$ & $4.2\times 10^{8}$ & $2.4\times 10^{8}$ & $1.9\times 10^{12}$ & $9.9\times 10^{11}$ & $7.9\times 10^{11}$ \\  \hline
\end{tabular}
\caption{Summary statistics for the total daily volume of (Panel A) market orders and (Panel B) limit orders. All volumes are in units of the counter currency.}
\label{tab:TradeSummaryStats}
\end{table}

The Hotspot FX data has several features that are particularly important for our study. First, the data lists all limit order arrivals and departures, irrespective of each order's ownership, so we can determine the complete set of all limit orders (irrespective of their owners' CCLs) for a given currency pair at any time during the sample period. By doing this at the time of each trade, we are able to calculate detailed statistics about the impact of CCLs on trade prices. Second, the small tick sizes on Hotspot FX enable us to observe market participants' price preferences (i.e., the prices at which they place orders) with a high level of detail. Data from platforms with larger tick sizes (such as Reuters and EBS) provide a more coarse-grained view that makes results more difficult to interpret, particularly for trades for which CCLs exert a small influence. Third, all limit orders on Hotspot FX represent actual trading opportunities that were available in the market. This is not the case on some other platforms, which allow institutions to post indicative quotes that do not constitute a firm commitment to trade.

For the purposes of our investigation, the Hotspot FX data also has some limitations. First, the data does not provide any way to identify financial institutions, nor does it allow us to ascertain which institutions participated in which trades. Therefore, our statistical analysis is limited to studying aggregate behavior across all institutions, rather than more detailed conditional 
properties. Second, the data does not include information about hidden orders. In the absence of further details about these orders, we exclude them from our study. Third, in some extremely busy periods, several limit order departures can occur at the same price in very rapid succession. 
Therefore, for some trades, it is not possible to determine exactly which limit order departure corresponds to a given trade. For each such trade, we use the limit order departure whose time stamp is closest to the reported trade time. We regard any incorrect associations from this approach as a source of noise in the data. To ensure that this choice does not influence our conclusions, we repeated all of our calculations when excluding all trades for which it is not possible to associate exactly one limit order departure, and we obtained results that are qualitatively the same as those that we report throughout the paper. We discussion all these points further in \citet{Gould:2016quasi}.

If a market order matches to several different limit orders, each partial matching appears as a separate line in the trade-data file, with a time stamp that differs from the previous line by at most $1$ millisecond. We regard all entries that correspond to a trade of the same direction and that arrive within $1$ millisecond of each other as originating from the same market order, and we record the corresponding statistics for this market order only once. For trades that match at several different prices (i.e., they ``walk up the book''), we record the volume-weighted average price (VWAP) as the price for the whole trade, and we calculate the corresponding skipping cost using this VWAP price.\footnote{Because each partial matching of a single market order is subject to the same CCLs, we regard it as inappropriate to study each such partial matching as a separate event, as doing so would produce long sequences of correlated data points from single market orders.}

Although the Hotspot FX data does not include information about market activity on Reuters or EBS, we do not regard this to be an important limitation of the present study. Due to the greater heterogeneity among member institutions on Hotspot FX than on Reuters or EBS (see Section~\ref{sec:ccls}), it seems reasonable to expect that CCLs have a larger impact on trade prices on Hotspot FX than they do on these other platforms. For example, large banks that trade on Hotspot FX may be unwilling to trade with small counterparties, and they may therefore assign them a CCL of $0$. By contrast, the CCLs between institutions on Reuters and EBS are likely to be much higher to reflect the confidence in large trading counterparties in the interbank market. By studying data from Hotspot FX, we are able to assess the impact of CCLs in a large and heterogeneous population.


\section*{Appendix B: Measuring Realized Volatility}\label{app:realized_variance}

In this appendix, we describe our methodology for measuring the realized volatility of the quote-price and trade-price series in our model. For a detailed discussion of this approach and its empirical performance, see \citet{Liu:2015does}.

For concreteness, we describe our methodology for buyer-initiated trades; we treat seller-initiated trades in an analogous way. For a given simulation of our model, let $X$ denote the total number of buyer-initiated trades that occur, let $A_1, A_2, \ldots, A_X$ denote the prices of these trades, and let $a_1, a_2, \ldots, a_X$ denote the ask-prices immediately before the arrival of these trades. For a given number $K$ of intervals and a given number $L$ of subsamples, let $T := X/K$ denote the sample width and let $\tau := T/L$ denote the subsample width. For a given lag $j$, we calculate the \emph{sell-side trade returns}
\begin{equation}\label{eq:rA}
	r^A_i(j) = \log\left(A_{\left\lfloor(i+1)T+j\tau\right\rfloor}\right)-\log\left(A_{\left\lfloor iT+j\tau \right\rfloor}\right),\quad \ i\in \left\{1,\ldots,K-1\right\}\,,
\end{equation}
where $\left\lfloor x \right\rfloor$ denotes the largest integer less than or equal to $x$. We then calculate
\begin{equation}
	v_A(j):=\sum_{i=1}^{K-1} \left(r^A_i(j)\right)^2\,. 
\end{equation}	
We repeat this process for each $j = 0,1,\ldots,L-1$ and calculate the \emph{sell-side trade-price quadratic variation}
\begin{equation}\label{eq:vA}
	v_A = \frac{1}{L}\sum_{j=0}^{L-1}v_A(j)\,.
\end{equation}
We calculate the \emph{sell-side quote-price quadratic variation} $v_b$ similarly from $a_1, a_2, \ldots, a_X$.

To identify a suitable value of $K$, we create volatility signature plots and choose values of $K$ within a plateau (see \citet{Andersen:2000great}). Other values of $K$ in the same plateau produce results that are qualitatively the same as those that we report.


\bibliographystyle{plainnat}
\bibliography{dphilLOBbib-rev}

\end{document}